\documentclass[twoside,12pt]{article}
\usepackage{epsfig}
\usepackage[utf8]{inputenc}
\usepackage[usenames]{color}

\def\NPA{{\em Nucl. Phys.} A}

\def\PTP{{\em Prog. Theor. Phys.}}

\def\PPNP{{\em Prog. Part. Nucl. Phys.}}
\def\NPB{{\em Nucl. Phys.} B}

\def\PLB{{\em Phys. Lett.} B}

\def\PRL{{\em Phys. Rev. Lett.}}
\def\PREV{{\em Phys. Rev.}}

\def\PRD{{\em Phys. Rev.} D}
\def\PRC{{\em Phys. Rev.} C}

\def\ZPA{{\em Z. Phys.} A}
\def\EPJA{{\em Eur. Phys. J.} A}
\def\EPJC{{\em Eur. Phys. J.} C}

\newcommand{\be}{\begin{equation}}
\newcommand{\ee}{\end{equation}}
\newcommand{\bea}{\begin{eqnarray}}
\newcommand{\eea}{\end{eqnarray}}

\begin{document}

\title{ \vspace{1cm} Dibaryons: Molecular versus Compact Hexaquarks} 
\author{H.\ Clement and T.\ Skorodko \\ \\
Physikalisches Institut der Universit\"at T\"ubingen \\
and\\
Kepler Center for Astro and Particle Physics, University of T\"ubingen,\\
Auf der Morgenstelle 14, D-72076 T\"ubingen, Germany\\emal: heinz.clement@uni-tuebingen.de\\
\\
} 
\maketitle

\begin{abstract} Hexaquarks constitute a natural extension of complex quark
  systems like also tetra- and pentaquarks do. To this end the current status of
  $d^*(2380)$ in both experiment and theory is reviewed. Recent
high-precision measurements in the nucleon-nucleon channel and analyses
thereof have established $d^*(2380)$ as an indisputable resonance in the
long-sought dibaryon channel. Important features of this $I(J^P) = 0(3^+)$
state are its narrow width and its deep binding relative to the
$\Delta(1232)\Delta(1232)$ threshold. Its decay branchings favor theoretical
calculations predicting a compact hexaquark nature of this state.   
We review the current status of experimental and theoretical studies on
$d^*(2380)$  as well as new physics aspects it may bring in the future. In
addition, we review the situation at the 
$\Delta(1232) N$ and $N^*(1440)N$ thresholds, where evidence for a number of
resonances  of presumably molecular nature have been found --- similar to the
situation in charmed and beauty sectors. Finally we briefly discuss the
situation of dibaryon searches in the flavored quark sectors. 
\end{abstract}
\eject
\tableofcontents


\section{Introduction}
\label{intro}

The recent observations of exotic multi-quark states in form of tetra- and
pentaquark systems in charmed and beauty meson and baryon sectors,
respectively, demonstrate that there exist more complex configurations in
nature than just quark-antiquark and three-quark systems --- as already
suggested by Gell-Mann in his pioneering presentation of the quark model
\cite{GellMann}. A natural 
extension of complexity is the quest for hexaquark systems, which provides the
transition to two-baryon, {\it i.e.} dibaryon systems.

Generally dibaryons are solely defined by their baryon quantum number $B =$
2. In this sense we know since 1932, when the deuteron was discovered
\cite{Urey} that at least a single one does exist. Due to its very small
binding energy of only 2.2 MeV the deuteron constitutes a large extended
hadronic molecule with a charge radius of 2.1 fm \cite{ADANDT,CODATA}. {\it
  I.e.}, the proton and the neutron inside the deuteron are on
average 4 fm apart from each other and do not overlap. 

All the time since then it has been questioned, whether there are more states
in the two-baryon system than just the deuteron groundstate with $I(J^P) =
0(1^+)$. Follow-up nucleon-nucleon ($NN$) scattering experiments revealed this
state in the $^3S_1$ partial wave to be the only bound state in the $NN$
system. 
Its isovector counterpart, the virtual $I(J^P) = 1(0^+)$ state in the
$^1S_0$ partial wave, was found to be already slightly unbound.  

With the recognition of quarks being the basic building blocks of hadrons, the
idea of dibaryons being not just hadronic molecules but rather clusters
  ("sixpacks") of quarks sitting in a common quark bag 
stimulated the dibaryon search enormously. A manyfold of quark models
predicting a huge number of dibaryon states initiated a rush of experimental
searches for such objects. Unfortunately, practically none of the many
claims for experimental evidences survived rigorous experimental checks. For a
review of the history of dibaryon predictions and searches see, {\it e.g.}
Ref. \cite{dibaryonreview}. 

This situation has changed about 10 years ago, when the CELSIUS/WASA 
\cite{prl2009} and the WASA-at-COSY collaborations \cite{prl2011,isofus}
started to report their 
experimental results obtained in a series of experiments on two-pion
production in $NN$ collisions and in neutron-proton scattering. In all
relevant two-pion channels the Lorentzian energy dependence of a narrow
isoscalar resonance -- named $d^*(2380)$ --- 
was observed. By measurement of polarized proton-neutron scattering and
its inclusion in the phase-shift analysis a circular counter-clockwise
move in the $^3D_3$ 
partial wave  was revealed establishing a pole with $I(J^P)=0(3^+)$ at
around 2380 MeV \cite{prl2014,RWnew}. From these investigations the branching
ratios of $d^*(2380)$ were determined for all its hadronic decays
\cite{BR}.   

In the following chapters 2 and 3  a short review is given
starting from the first solid 
observation of $d^*(2380)$ in the double-pionic fusion measurements at
CELSIUS/WASA \cite{prl2009} and WASA-at-COSY \cite{prl2011} until its current
status in hadronic and electromagnetic excitation and decay processes. In
chapter 4 the status of theoretical work on $d^*(2380)$ is reviewed with
emphasis on the width issue and the key question, whether it constitutes a
compact hexaquark or a dilute molecular system.

Chapter 5 deals with resonance structures
at $\Delta(1232)N$ and $N^*(1440)N$ thresholds pointing to dibaryonic states
of molecular character --- in analogy to the situation for tetra- and
pentaquark systems in charm and beauty sectors. Finally, the current dibaryon
situation in flavored quark sectors is shortly touched in chapter 6.

\section{Pion Production in Nucleon-Nucleon Collisions and the Issue of Resonances}
\label{sec-pion}

Resonances in single- and two-baryon systems decay preferentially by
emission of one or several pions. Hence  pion production in $NN$
collisions gives access to the physics of resonances both in baryon and
dibaryon systems. The latter are of particular interest here. The oldest
prediction of six-quark objects decaying by pion emission dates back to Dyson
and Xuong \cite{Dyson}, who --- based on SU(6) symmetry considerations ---
predicted the existence of six non-strange dibaryon states. 
   
Since there existed no detailed data base on pion production in $NN$ collisions,
a systematic study -- in particular of two-pion production -- started in the
nineties at CELSIUS and was continued later-on at COSY using the hermetic WASA
detector. All CELSIUS/WASA and WASA-at-COSY measurements on single- and
multiple-pion production reported here were carried out exclusively and
kinematically complete --- in most cases kinematically over-constrained, in
order to improve the momentum resolution by kinematic fits and in order to
provide data free of background. 

\subsection{\it Single-Pion Production --- Early Results on Dibaryonic States
  near the $\Delta(1232)N$ Threshold}
\label{sec-singlepion}

The search for resonances in the system of two baryons dates back to
the fifties, when first measurements of the $\pi d \to pp$ reaction at
Dubna \cite{Neganov,Neganov1,Mesh} indicated a resonance-like structure near
the $\Delta N$ 
threshold connected to the $^1D_2$ partial wave in the $NN$ system.
Later-on high quality data on total and differential cross sections
and polarization observables for $pp$ and $\pi d$ elastic scattering as well
as $\pi d \to pp$ and $pp \to \pi d$ reactions revealed a pronounced looping of
the $^1D_2$ $NN$ partial wave in the Argand diagram representing a pole of a
resonance with $I(J^P) = 1(2^+)$, mass $m \approx$ 2148 MeV and width $\Gamma
\approx 120 MeV$ \cite{FA91,SAID97}.

Though the clear looping is in favor of a true s-channel resonance,
the close neighborhood of its mass to that of the $\Delta N$ threshold and
the compatibility of its width with that of $\Delta(1232)$ casted doubts
on its s-channel nature. It has been argued that the observed features
could be merely a threshold phenomenon and the observed looping just a
reflection of the usual $\Delta$ excitation process in the presence of
the other nucleon, which due to the threshold condition has to be at
rest relative to the active one.

The situation about this resonance structure has been discussed
controversial in many papers, see, {\it e.g.} \cite{dibaryonreview}. In a
number of 
publications Hoshizaki demonstrated that this resonance structure
constitutes a true S-matrix pole rather than a threshold cusp or a
virtual $\Delta N$ state\cite{hos1,hos2}. Similar conclusions were reached by
Ueda {\it et al.} \cite {Ueda1}.

The resonance structure seen in the $^1D_2$ $NN$-partial wave is the by far
most pronounced one seen in $NN$ scattering and $\pi d \rightleftharpoons NN$
reactions. 
But also in other partial waves a resonant behavior had been noted in
the region of the $\Delta N$ threshold, though by far not as spectacular.
In partial wave solutions of the SAID analysis group, {\it e.g.}, also the
$^3P_2-$$^3F_2$, $^3F_3$ and $^3F_4-$$^3H_4$ $NN$-partial waves exhibit a clear
looping in the Argand diagram \cite{FA91,SAID97,Arndt}.

The fact that all these states near the $\Delta N$ threshold exhibit a
width close to that of the $\Delta(1232)$ is not too surprising, since
the available phase space of a $\Delta N$ state for a fall-apart decay
into its components $N$ and $\Delta$ is tiny close to the $\Delta N$ threshold
and hence the only sizeable decay contribution arises from the decay
of the component $\Delta$. We will return to the discussion of states near
thresholds in chapter 5. In the next chapters first the situation about
the hitherto only example of a deeply bound (relative to the $\Delta\Delta$
threshold) dibaryon state, the $d^*(2380)$, will be reviewed.

\subsection{\it Two-Pion Production --- Observation of the Deeply Bound
$\Delta(1232)\Delta(1232)$ State $d^*(2380)$}
\label{sec-twopion}

\subsubsection{\it $pp$-induced two-pion production}
\label{sec-pptwopion}

The two-pion
production program at CELSIUS started out in 1993 with exclusive and
kinematically complete high-statistics measurements of $pp$-induced two-pion
production from the threshold region up to the GeV region. 

As a result of these
systematic studies it was found that isovector induced two-pion production up
to $\sqrt s \approx$ 3 GeV is well described by the
conventional process of $t$-channel meson exchange leading to the excitation
of the $N^*(1440)$ Roper resonance and the excitation of the
$\Delta(1232)\Delta(1232)$ system. Whereas the first process dominates at
lower beam energies close to threshold, the latter dominates at energies above
1 GeV, {\it i.e.} $\sqrt s >$ 2.4 GeV.  

\begin{figure} 
\centering
\includegraphics[width=8cm,clip]{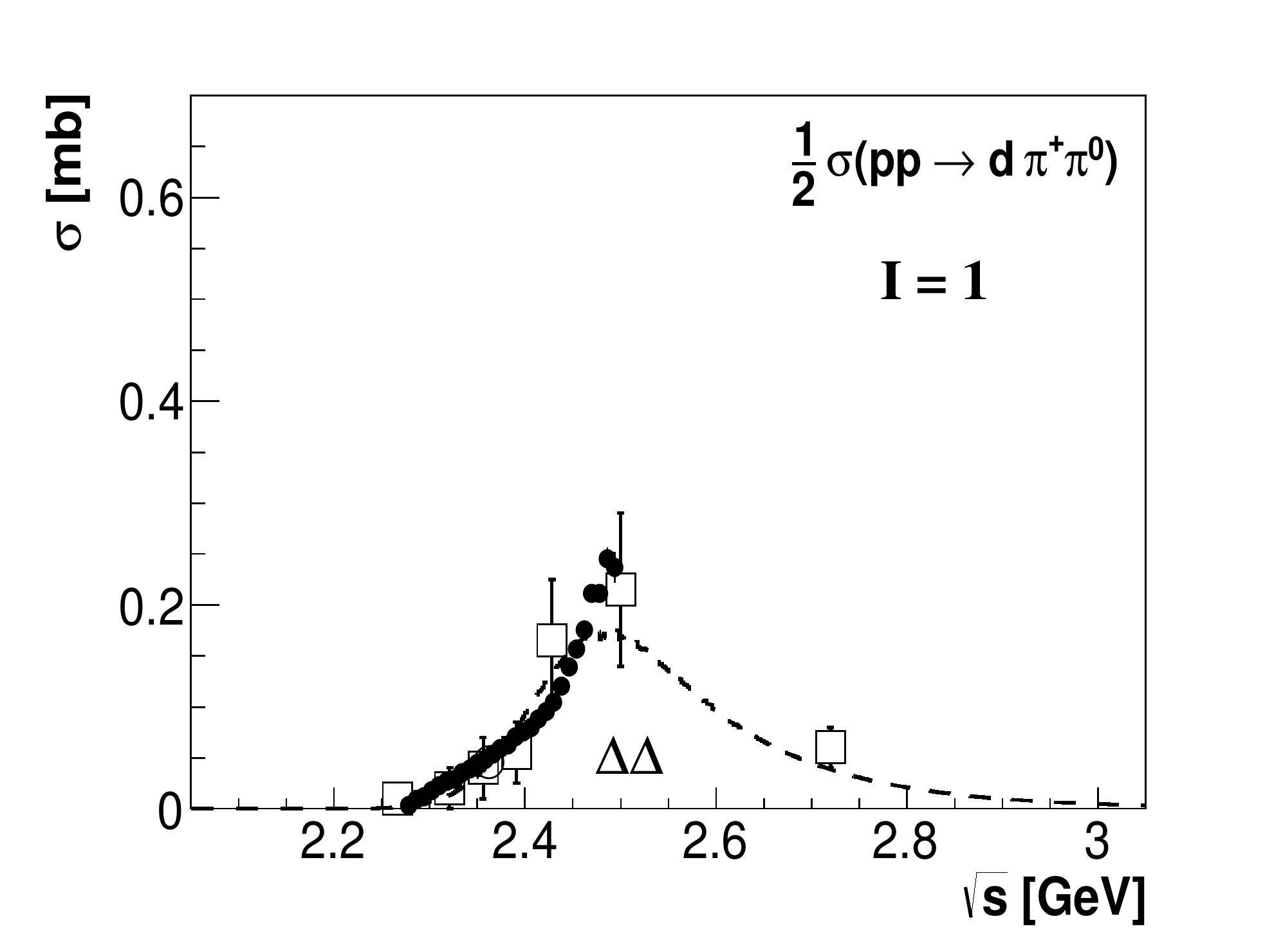}\\
\includegraphics[width=8cm,clip]{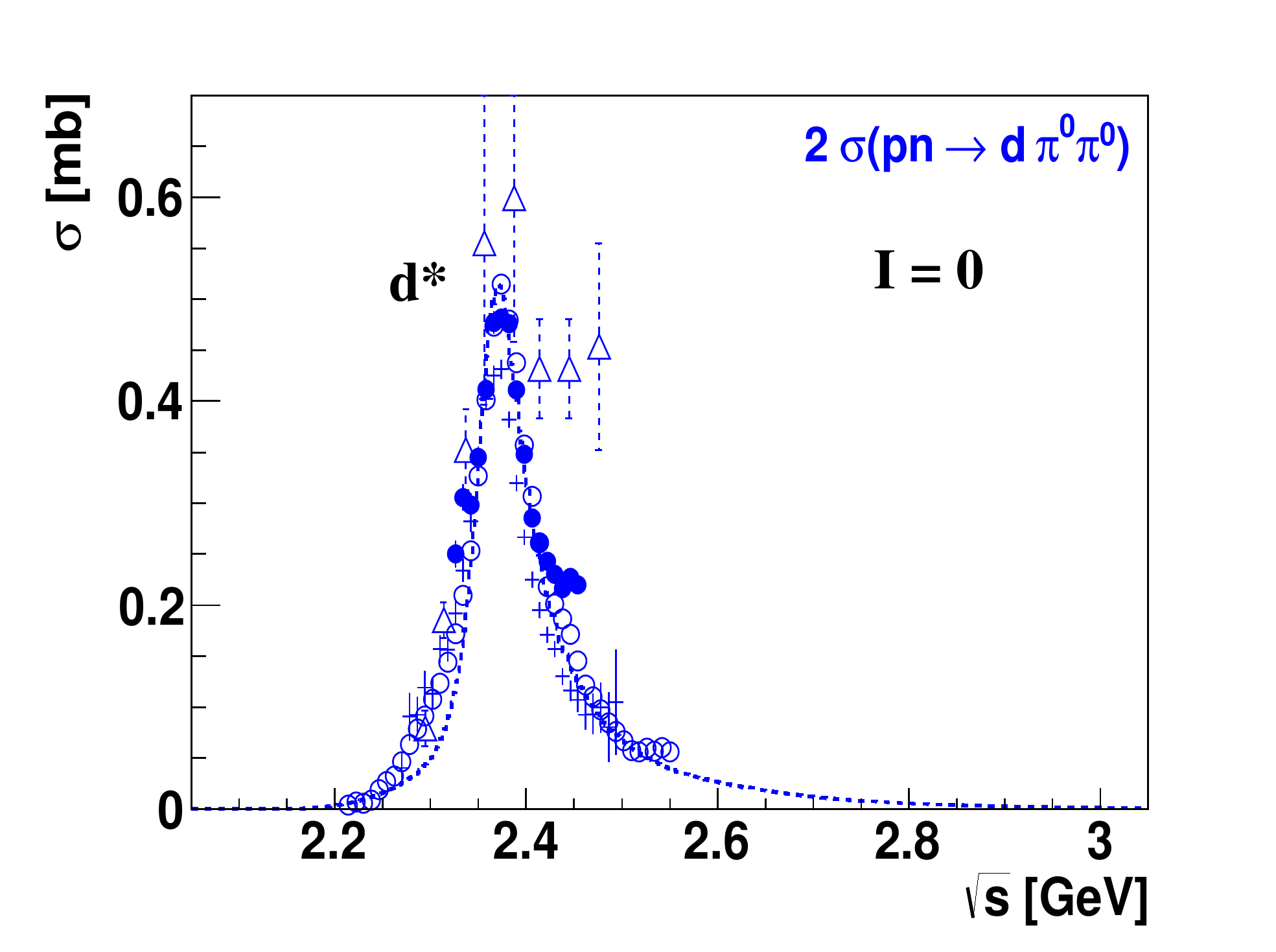}\\
\includegraphics[width=8cm,clip]{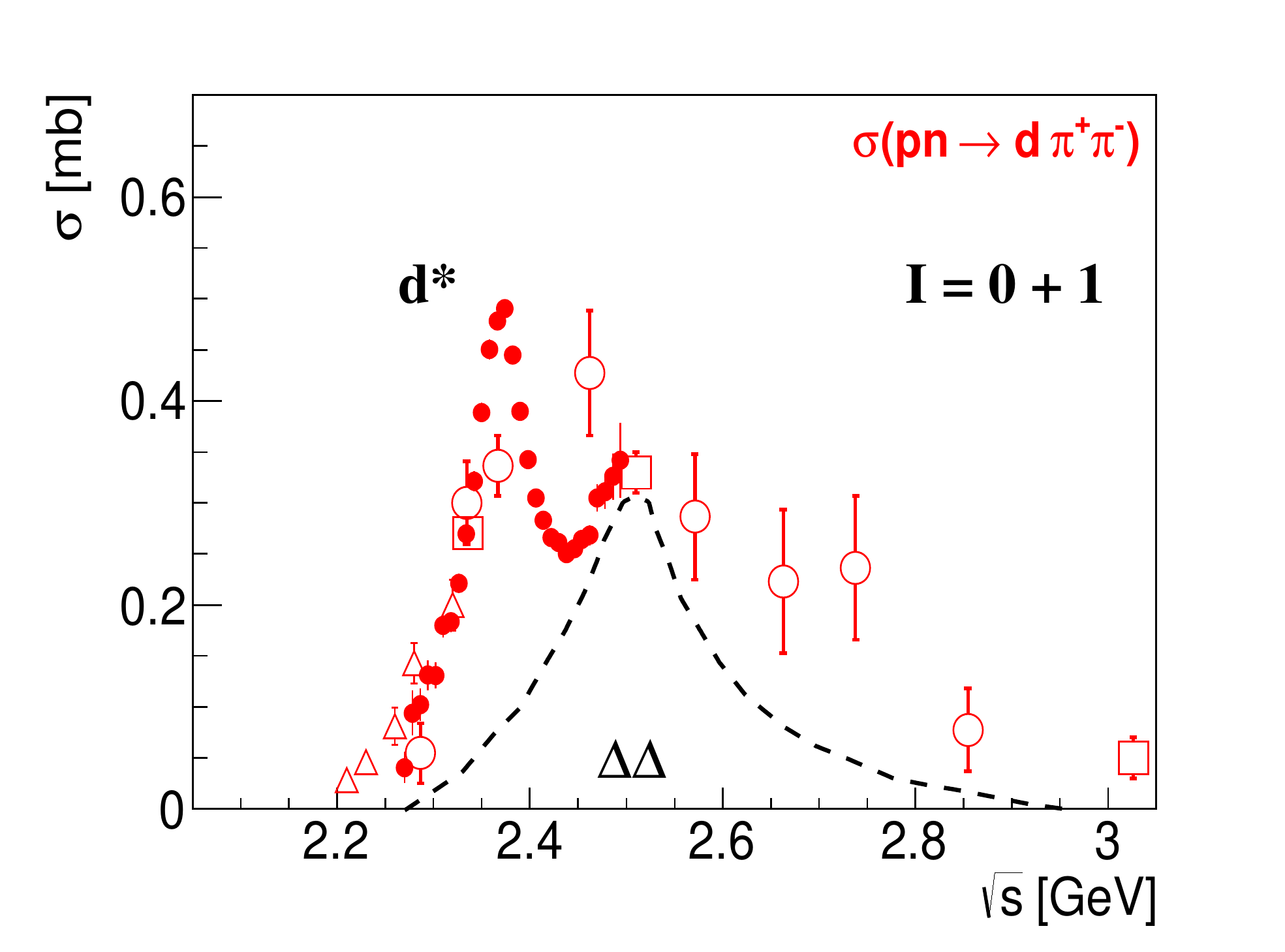}
\caption{
Total cross section of the double-pionic fusion to deuterium and
its isospin decomposition.
Top panel: the isovector part of the $pn \to d\pi^+\pi^-$ reaction given by
half the cross section of the $pp \to d\pi^+\pi^0$ reaction. Solid dots
show WASA-at-COSY \cite{isofus} results, open symbols previous results 
\cite{FK,Bystricky,Shimizu}. The dashed curve represents a $t$-channel
$\Delta\Delta$ calculation fitted in height to the data \cite{FK}.  
Middle panel: The isoscalar part of the $pn \to
d\pi^+\pi^-$ reaction given by twice the cross section of the $pn \to
d\pi^0\pi^0$ reaction. The CELSIUS/WASA results \cite{prl2009} are shown by
open triangles. The other symbols refer to WASA-at-COSY measurements
\cite{prl2011,isofus}. The dotted line denotes the $d^*$ resonance curve
with momentum dependent widths \cite{abc}, mass m = 2370 MeV
and total width $\Gamma$ = 70 MeV. 
Bottom panel: the isospin-mixed reaction $pn \to
d\pi^+\pi^-$. Solid dots represent WASA-at-COSY measurements, open symbols
previous bubble-chamber measurements at DESY (circles) \cite{desy}, Dubna
(squares) \cite{dubna} and Gatchina (triangles) \cite{Dakhno}. The dashed line
represents the $t$-channel $\Delta\Delta$ excitation.
From Ref. \cite{isofus}.
}
\label{fig-isofus}       
\end{figure}

This conclusion includes also the
isovector double-pionic fusion process $pp \to d\pi^+\pi^0$. Measurements of
its differential cross sections in the region $\sqrt s \approx$ 2.4 GeV are in
good agreement with $t$-channel $\Delta\Delta$ calculations.
And the energy
dependence of its total cross section exhibits a broad resonance-like
structure with a width of about 2$\Gamma_\Delta$ in accord with the
$t$-channel $\Delta\Delta$ calculations \cite
{isofus,FK} --- see top panel of Fig.~\ref{fig-isofus}. 

\subsection{\it $pn$-induced double-pionic fusion: observation of a narrow resonance}
\label{sec-nptwopion}

When $pn$-induced two-pion production was looked at, the situation changed
strikingly. The measurements were carried out with either a deuteron
beam or deuterium target by taking advantage of the quasi-free process, {\it
  e.g.}, by looking on the process $pd \to d\pi^0\pi^0 + p_{spectator}$ within
the $pd \to dp\pi^0\pi^0$ reaction. Since all these measurements were
exclusively and kinematically complete (in most cases even over-constrained
allowing kinematic fitting with improving thus resolution and purity of the
collected events), the effective total energy of the $pn \to d\pi^0\pi^0$
subprocess was known on a event-by-event basis. That way the energy dependence
of the quasi-free process could be measured over an appropriate energy range
with a single beam energy setting.     

For the $d\pi^0\pi^0$ channel there were no previous measurements at all,
since a hermetic detector like WASA with its capability to detect both charged
and uncharged particles over a solid angle of nearly $4\pi$ was not available
at other installations for hadron research.

By use of isospin relations \cite{Bystricky,Dakhno} and isospin
recoupling in case of an intermediate $\Delta\Delta$ system
\cite{dibaryonreview}, the cross section of the $t$-channel $\Delta\Delta$
process in the $d\pi^0\pi^0$ channel can be determined 
to be only 1/5 of that in the $d\pi^+\pi^0$ channel, {\it i.e.}, about 0.04 mb
at $\sqrt s \approx$ 2.5 GeV, where the $t$-channel $\Delta\Delta$
process peaks. Due to this low cross section of conventional processes, this
reaction channel is predestinated for the observation of unconventional
isoscalar processes, so to speak the "golden" channel. 

\begin{figure} 
\centering
\includegraphics[width=8.5cm,clip]{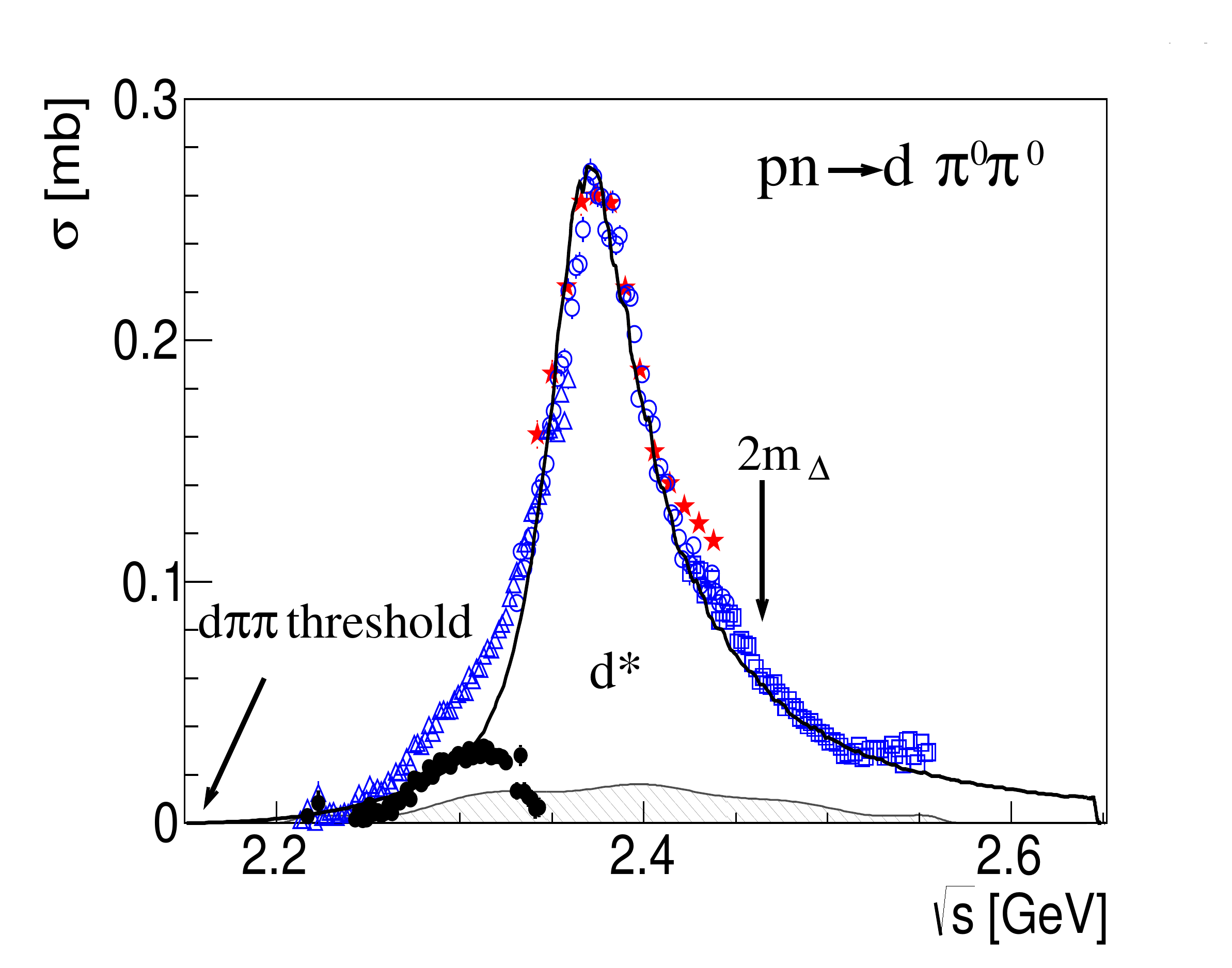}\\
\includegraphics[width=6.5cm,clip]{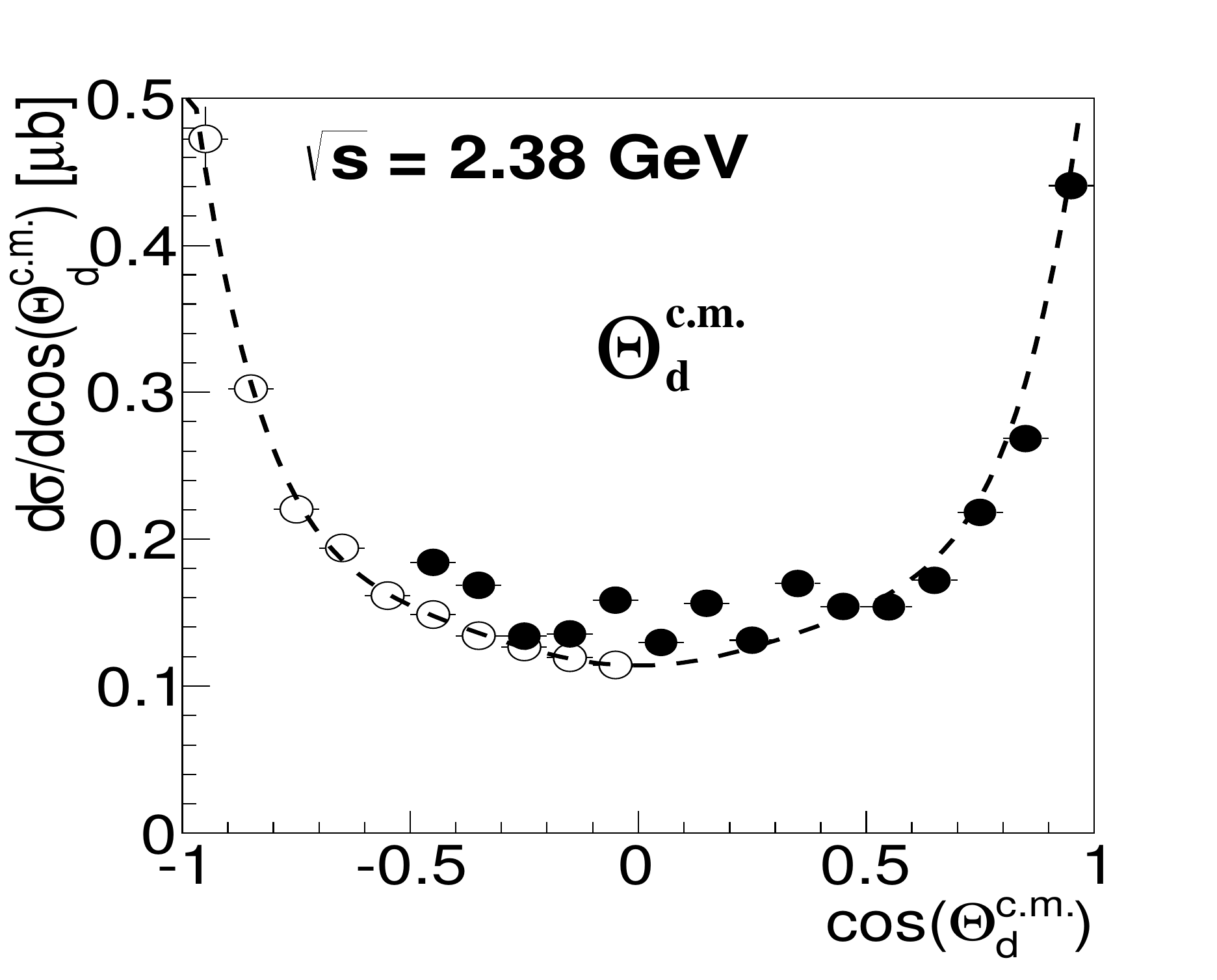}
\includegraphics[width=6.5cm,clip]{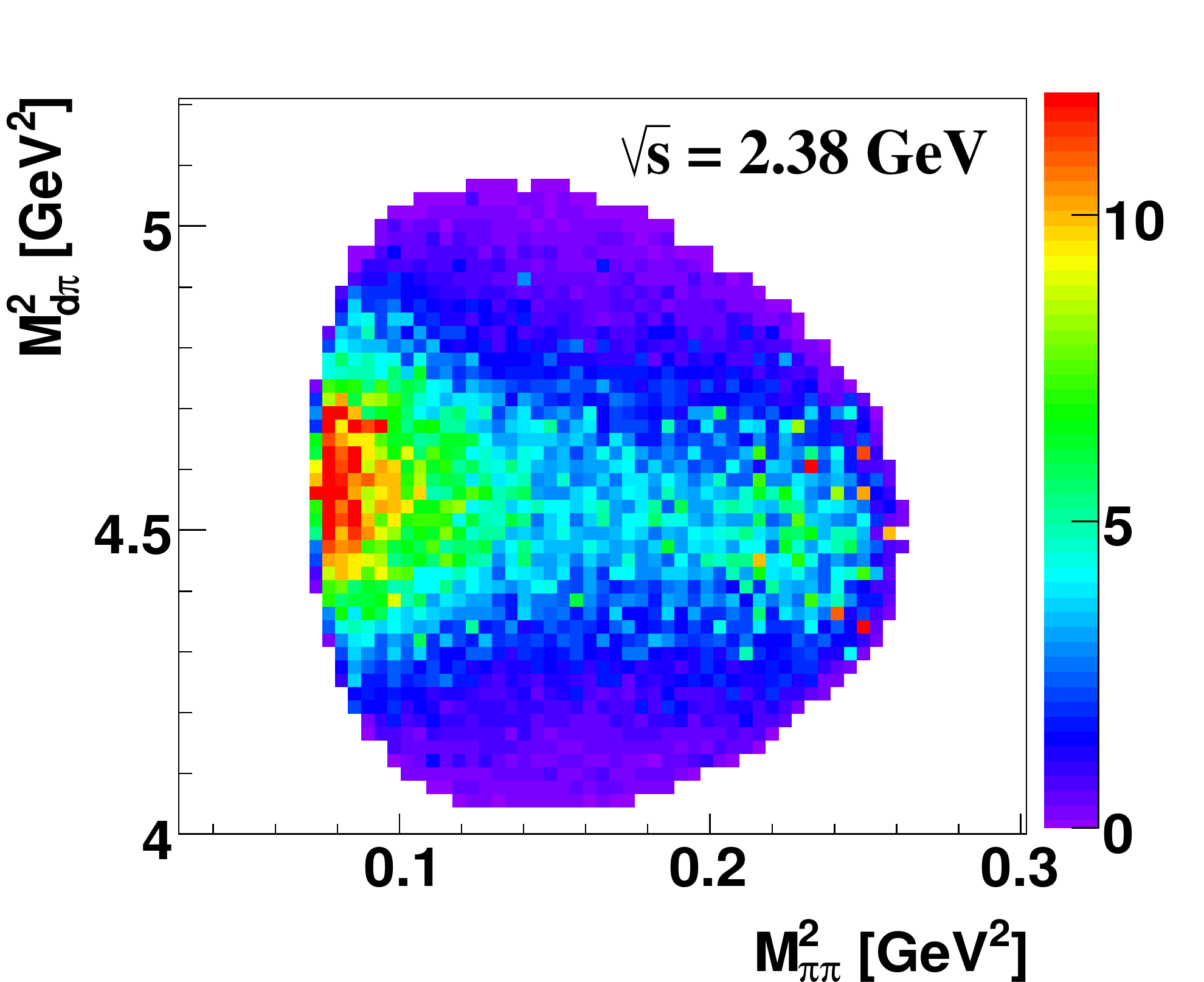}
\includegraphics[width=6.5cm,clip]{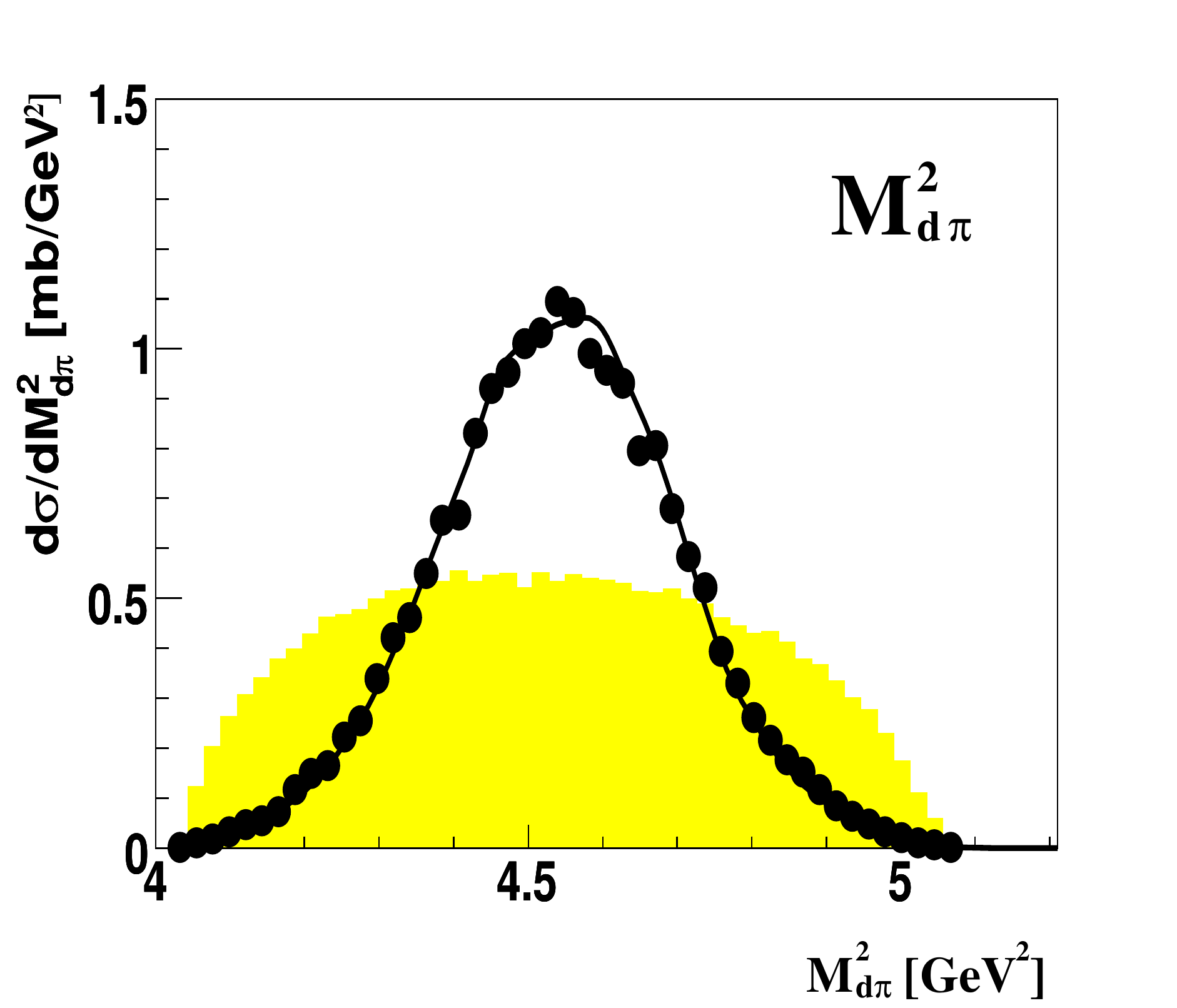}
\includegraphics[width=6.5cm,clip]{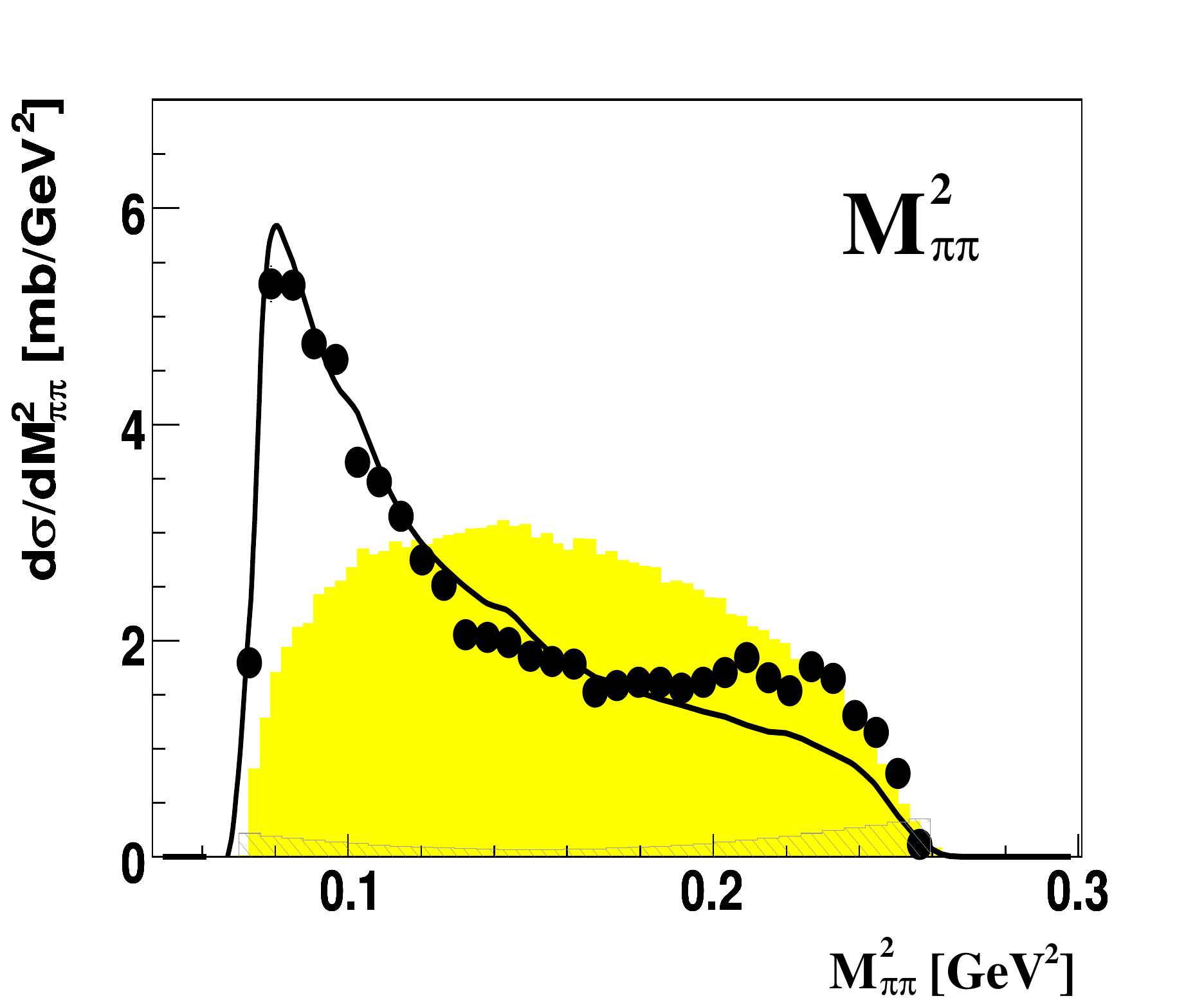}
\caption{Measurements of the "golden" reaction channel $pn \to d\pi^0\pi^0$
  with WASA at COSY. 
  Top: total cross section exhibiting the pronounced
  resonance structure. The blue open symbols show the data of
  Ref.~\cite{prl2011}. They have been normalized in absolute scale to the data
  of Ref.~\cite{isofus}, which are plotted by red stars. The black shaded area
  represents an estimate of systematic uncertainties. The solid curve shows
  a calculation of the $d^*(2380)$ resonance with momentum-dependent widths
  according to Ref.~\cite{abc} and including $t$-channel Roper and
  $\Delta\Delta$ excitations as background reactions. The filled circles
  represent the difference 
  between this calculation and the data in the low-energy tail of $d^*(2380)$.
  Middle: deuteron angular distribution (left) and Dalitz plot (right) at the
  peak energy of $\sqrt s$ = 2.38 GeV. Open and solid circles refer to
  measurements with the spectator proton in the target and in the beam
  (reversed kinematics), respectively. The dashed curve gives a Legendre fit
  with $L_{max}$ = 6 corresponding to J~=~3.
  Bottom: Dalitz plot projections yielding the distributions of the
  squares of the $d\pi^0$- (left) and $\pi^0\pi^0$- (right) invariant
  masses. The low-mass enhancement in the latter spectrum denotes the so-called
  ABC effect. The solid lines represent a calculation of the
  $pn \to d^*(2380) \to \Delta^+\Delta^0 \to d\pi^0\pi^0$ process. From
  Refs. \cite{prl2011,Catania2014,poldpi0pi0}.
}
\label{fig-dpi0pi0}       
\end{figure}

The measurements for this channel \cite{prl2009,prl2011,isofus} displayed in
the middle panel of Fig.~\ref{fig-isofus} as well as in
Fig.~\ref{fig-dpi0pi0}, indeed revealed the cross section around 
2.5 GeV to be of this magnitude. However, the big surprise was that at lower
energies a 
much larger cross section was observed exhibiting a pronounced narrow
resonance-like structure, which can be very well fitted by a Breit-Wigner
ansatz with momentum dependent widths \cite{abc},
 mass m = 2370 MeV and total width $\Gamma$ = 70 MeV -- see dotted line in the
middle panel of Fig.~\ref{fig-isofus}. 

The measurement of the deuteron angular distribution  displayed in
Fig.~\ref{fig-dpi0pi0} led to a $J$~=~3 assignment for the
resonance structure \cite{prl2011}. Together with the isoscalar character
of the  $pn \to d\pi^0\pi^0$ reaction this gives the isospin-spin-parity
combination $I(J^P) = 0(3^+)$. Due to its
isoscalar character and the baryon number $B =$2 the resonance structure is
formally compatible with an excited state of the deuteron, hence its
denotation as $d^*$. 

The Dalitz plot and its mass-squared projections are shown in
Fig.~\ref{fig-dpi0pi0}. Together with the $N\pi^0$ angular distribution
\cite{prl2011} they suggest a $\Delta\Delta$ configuration in relative
$s$-wave as an intermediate configuration, which according to the observed
mass of 2370 MeV must be bound by about 90 MeV relative to the nominal
$\Delta\Delta$ threshold mass of 2464 MeV \cite{prl2011}.

In measurements of the $pn \to d\pi^+\pi^-$ reaction (bottom panel
of Fig.~\ref{fig-isofus}) and the
isospin decomposition \cite{isofus,Bystricky,Dakhno} of its cross section
according to the relation

\begin{equation}
\sigma (pn \to d\pi^+\pi^-) = 2\sigma(pn \to d\pi^0\pi^0) + \frac1 2 \sigma(pp
\to d\pi^+\pi^0).
\end{equation}

it has been demonstrated that the resonance structure shows up only in the
isoscalar part of the double-pionic fusion to the
deuteron, but not in its isovector part, {\it i.e.} it has a definite isospin
$I$~=~0.

\subsubsection{\it The $d^* \to \Delta\Delta$ decay vertex: ABC effect}
\label{sec-d*ABC}

The pronounced low-mass enhancement observed in the Dalitz plot and its
projection onto the $\pi\pi$-invariant mass-squared as displayed
Fig.~\ref{fig-dpi0pi0} is very remarkable. In fact, such low-mass
enhancements had been noticed already before in double-pionic fusion
experiments. Actually, they laid the trace for the discovery of $d^*$ at WASA
\cite{dibaryonreview,MB}. 

In 1960 Abashian, Booth and Crowe \cite{ABC} noticed an enhancement in
the $^3$He missing mass spectrum of the inclusively
measured $pd \to ^3$HeX reaction. This enhancement occured just in the kinematic
region corresponding to the emission of two pions with low $\pi\pi$-invariant
mass. Follow-up measurements revealed this enhancement 
to occur in the double-pionic fusion reactions $pn \to d\pi\pi$, $pd \to
^3$He$\pi\pi$  and $dd \to ^4$He$\pi\pi$, but not in the fusion to $^3$H,
where an isovector pion pair is emitted.

In all the years since then no conclusive explanation could be presented for
the observed low-mass enhancement in spite of many theoretical attempts. Hence
it was just abbreviated as "ABC" effect in the literature using the
initials of the authors Abashian, Booth and Crowe, who noticed this
enhancement first.

The WASA measurements of the complete double-pionic fusion to the deuteron
comprising all three reactions $pp \to d\pi^+\pi^0$, $pn \to d\pi^0\pi^0$ and
$pn \to d\pi^+\pi^-$ deciphered now this effect to be stringently correlated
with the appearance of the isoscalar resonance structure $d^*$
\cite{prl2011,isofus,FK} in double-pionic fusion processes. There the ABC
effect just reflects the vertex function of the decay vertex $d^* \to 
\Delta\Delta$ and shows up in the $\pi\pi$ invariant mass spectrum only, if
the nucleons in the final state fuse to a bound system
\cite{abc}. Subsequent WASA experiments showed that also in the double-pionic
fusions to $^3$He and $^4$He the dibaryon resonance $d^*$ is formed, though it
appears much broadened there due to collision damping with the surrounding
nucleons \cite{dibaryonreview,3he,4he}.

\subsubsection{\it $d^*$ resonance structure in non-fusing isoscalar two-pion
  production}
\label{sec-d*twopion}

Recently also the non-fusion two-pion production reactions $pn \to
pp\pi^0\pi^-$ \cite{pp0-}, $pn \to pn\pi^0\pi^0$ \cite{np00}, $pn
\to pn \pi^+\pi^-$ \cite{np+-} have been investigated. All
these channels are isospin-mixed, {\it i.e.} contain both isoscalar and
isovector contributions.  Hence the $d^*$ signal appears just on
top of a substantial and --- due to its four-body character --- steeply rising
background of conventional processes. Nevertheless it  still shows up clearly
in the energy dependence of the total cross sections for these reaction
channels. 

By use of the isospin-decomposition of $NN$-induced two-pion production
\cite{dibaryonreview,Bystricky,Dakhno} the expected size of the $d^*$
contribution in these channels can be easily estimated.  A more detailed
treatment takes into account also the different phase-space situation, when
the deuteron is replaced by the unbound $pn$ system in these reactions
\cite{CW,AO}.

In summary, all $NN$-induced two-pion production channels are in accordance
with the appearance of an $I(J^P) = 0(3^+)$
dibaryon resonance at 2.37 GeV with a width of 70 MeV. Even in the channels,
which are only partially isoscalar, the $d^*$ contribution 
is still the dominating process. The conventional $t$-channel
processes there  underpredict the data in the $d^*$ energy region by factors
two to four \cite{pp0-,np00,np+-}.

\subsubsection{\it $d^*(2380)$ -- a resonance pole in $np$ scattering}
\label{sec-d*np}

If the resonance structure $d^*$ observed in two-pion production
indeed is a true $s$-channel resonance, then it has to show up in principle also
in the entrance channel, {\it i.e.} in the $np$ scattering channel. There it
has to produce a pole in the partial waves corresponding to $I(J^P) =
0(3^+)$, {\it i.e.} in the coupled partial-waves  $^3D_3$ - $^3G_3$. 

The expected resonance contribution to the elastic $np$ scattering can be
calculated from the know\-ledge of the resonance contributions to the various
two-pion production channels --- under the assumption that there is no decay
into the isoscalar single-pion production channel, which is forbidden to first
order in case of an intermediate $\Delta\Delta$ formation. In Ref. \cite{BR}
this resonance contribution has been estimated to be about 170~$\mu$b, a
value, which is tiny compared to the value of nearly 40 mb for the total $np$
cross section.  

The analyzing power angular distribution of the elastic scattering is 
a particularly suitable observable to sense such a small contribution 
of $d^*(2380)$, since it is
composed just of interference terms in the partial waves and hence sensitive
to even small terms in the coupled $^3D_3 - ^3G_3$ partial waves. In the
angular distribution of the 
analyzing power the contribution of a resonance with $J =$ 3 is given by the
angular dependence of the associated Legendre polynomial 
$P^1_3$. Therefore the resonance contribution is expected to be largest at
90$^\circ$, which is also the angle, where the differential cross
section is smallest. For the sensitivity of other observables to the $d^*$
resonance see Ref.~\cite{RWnew}.

\begin{figure} 
\centering
\includegraphics[width=10.5cm,clip]{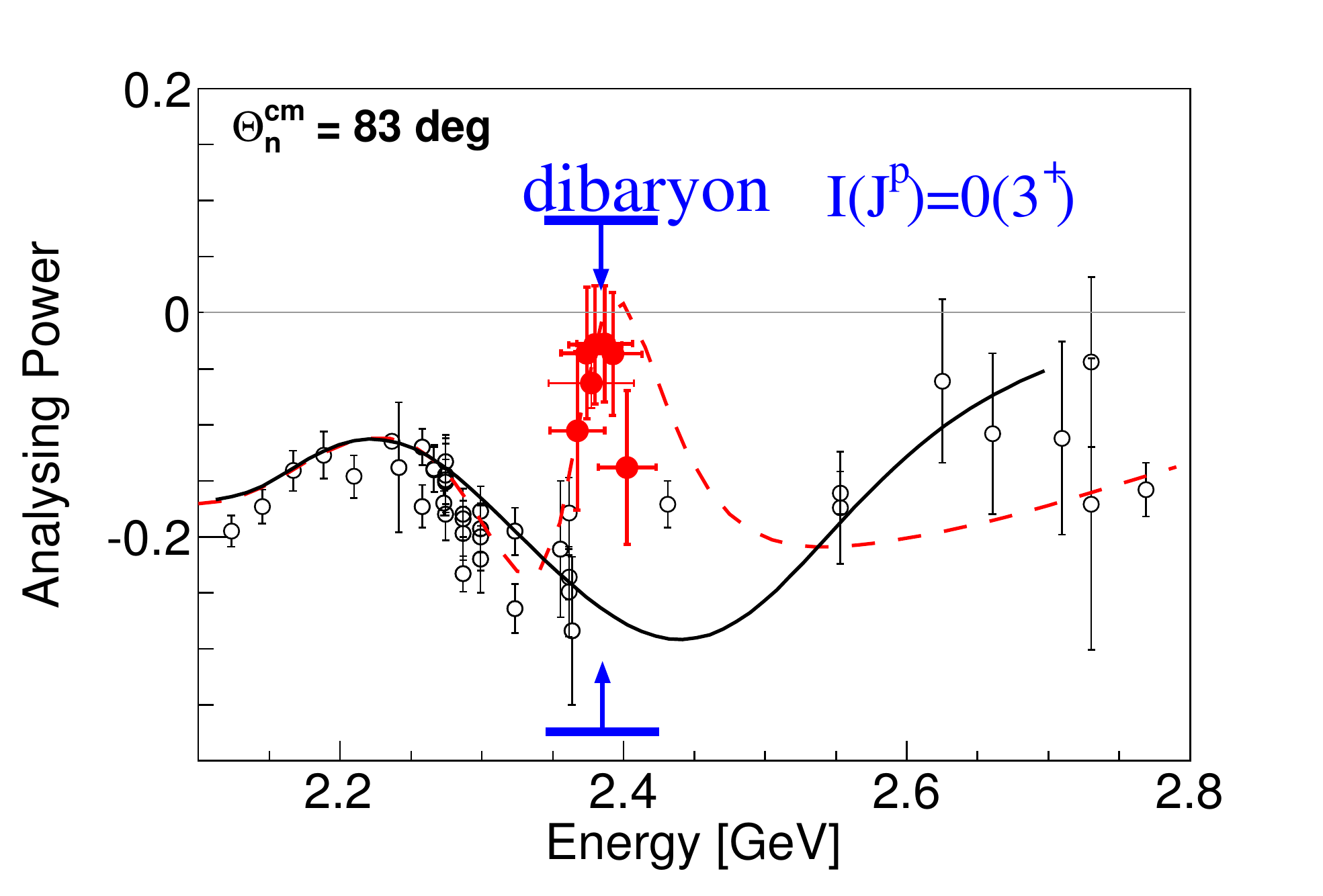}\\
\includegraphics[width=10.5cm,clip]{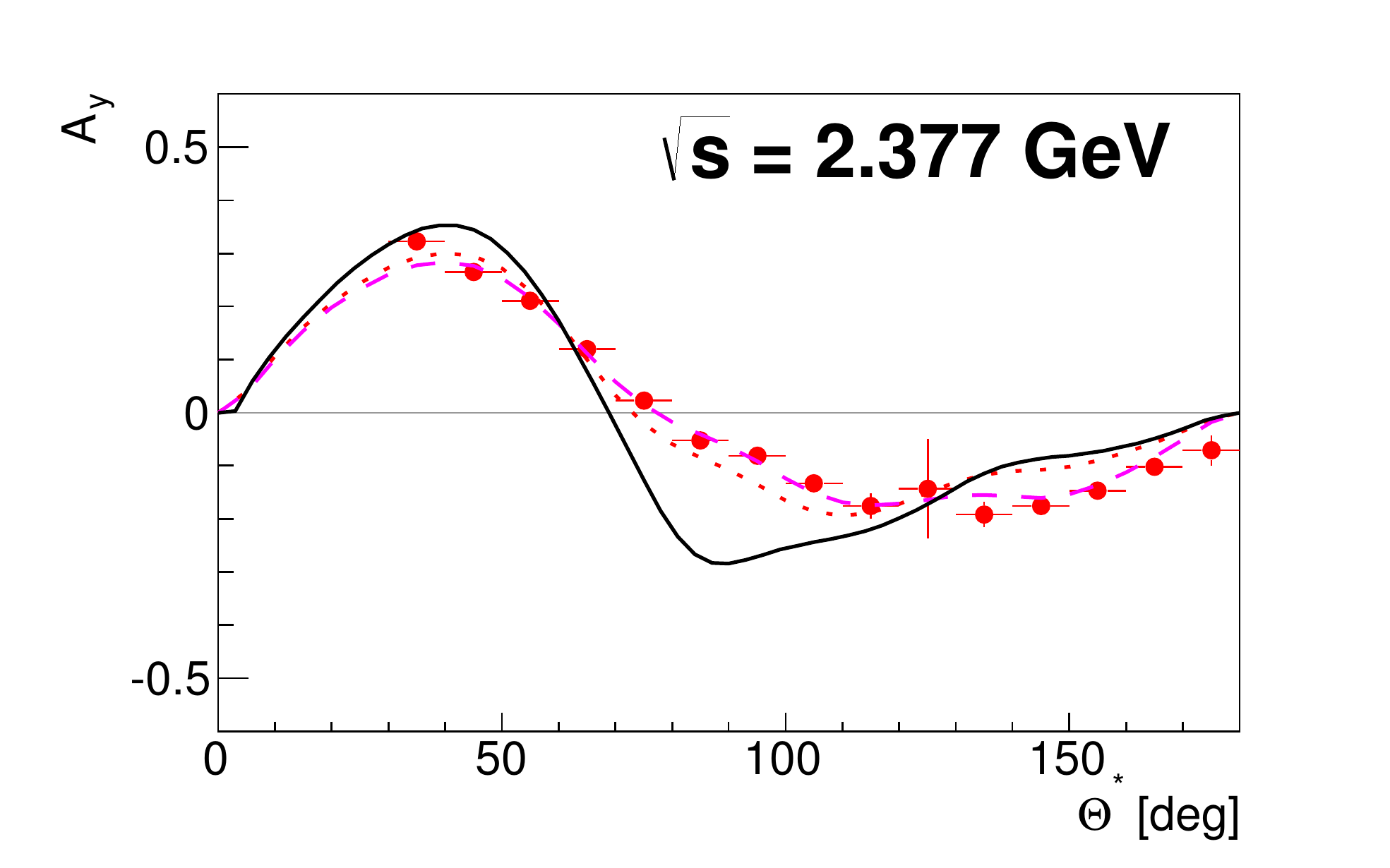}\\
\includegraphics[width=10.5cm,clip]{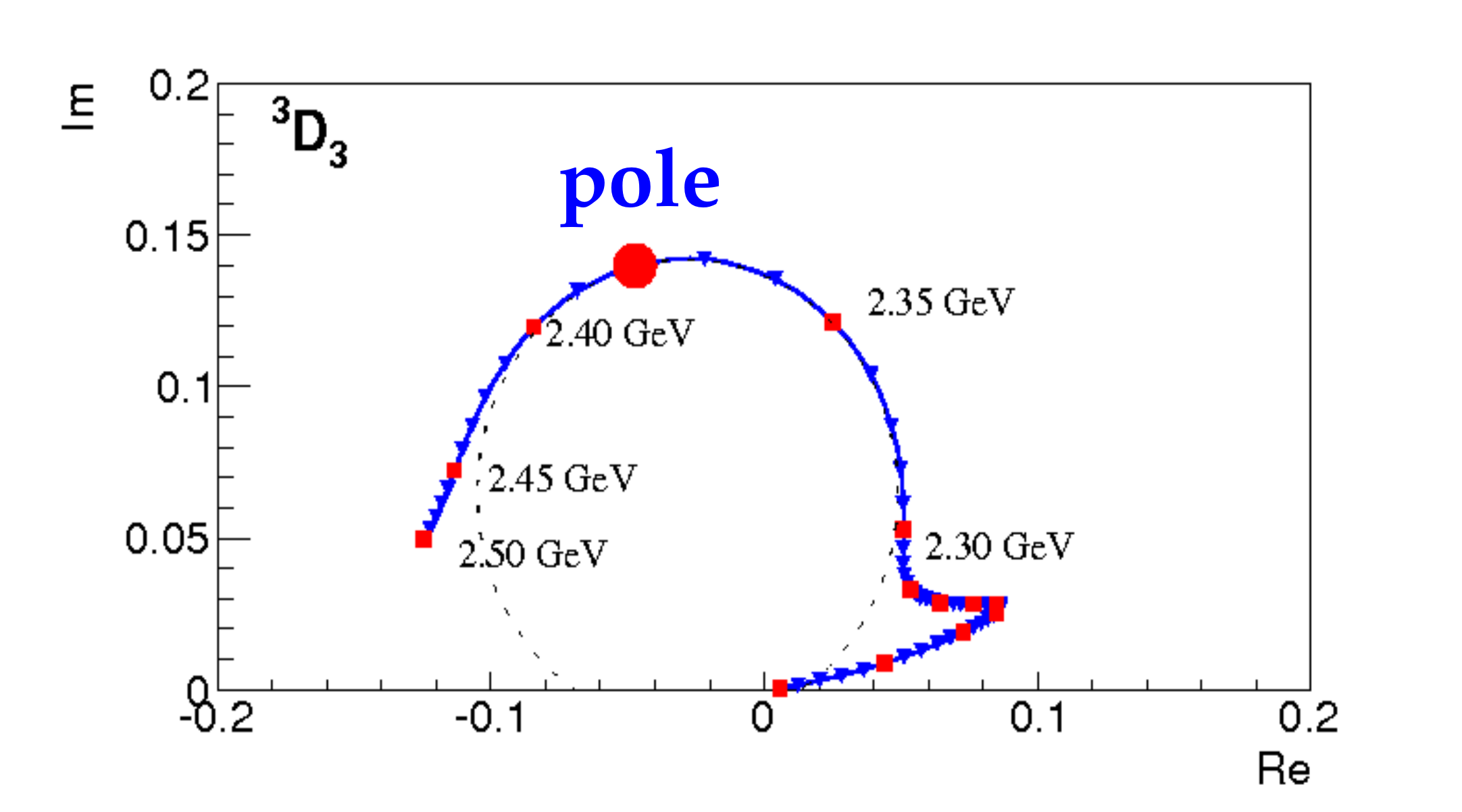}
\caption{Analyzing power data for elastic $pn$ scattering in the
  $d^*(2380)$ energy region and their partial-wave analysis
  \cite{prl2014,RWnew,npfull}.  Top: energy
  dependence of the analyzing power in the vicinity of $\Theta_n^{cm} =
  90^\circ$, where the effect of the $d^*(2380)$ resonance is expected to be
  largest. The solid circles denote WASA results, the open symbols previous
  data \cite{Ball,Lesquen,Makdisi,Newsom,Arnold,Ball1,McNaughton,Glass}. The solid line gives the previous SAID partial-wave solution,
  the dashed line the new SAID solution after including the WASA data. Middle: 
  Angular distribution of the analyzing power $A_y$ at the resonance
  energy. The curves have the same meaning as in the top figure. Bottom: Argand
  diagram of the new SAID solution for the $^3D_3$ partial wave with a pole at
  2380 MeV. The thick solid circle denotes the pole position. From
  Ref. \cite{dibaryonreview}.
}
\label{fig-pn}       
\end{figure}

 In the region of interest for the $d^*$ issue there existed no analyzing power
 data from previous measurements. Precise measurements at SACLAY ended just
 below the $d^*$ region 
 \cite{Ball,Lesquen}. Hence corresponding analyzing power measurements
 extending over practically the full 
angular range were undertaken with WASA at COSY --- again in the quasi-free
mode. By use of inverse kinematics a polarized deuteron beam was directed onto 
the hydrogen pellet target \cite{prl2014,npfull}.

The WASA data are shown in  Fig.~\ref{fig-pn} together with previous
measurements
\cite{Ball,Lesquen,Makdisi,Newsom,Arnold,Ball1,McNaughton,Glass}. The top
panel displays the energy dependence of the analyzing power near a
center-of-mass scattering angle of 90$^\circ$, where the  $d^*$ resonance
effect is expected to be largest. A pronounced narrow resonance-like
structure is observed in the data at the $d^*$ energy position. Accordingly,
the measured angular distribution of the analyzing power, displayed in the
middle panel  of Fig.~\ref{fig-pn}, deviates from the conventionally expected
distribution largest in the 90$^\circ$ region. In both panels the solid line
shows the solution SP07 from the SAID partial-wave analysis prior to the WASA
measurements \cite{SP07}.

The  subsequent partial-wave analysis  by the SAID group including now the
WASA data is given by the dashed and dotted lines, respectively in 
both top and middle panels. This partial-wave solution, denoted AD14, finds a
pole in the coupled $^3D_3 - ^3G_3$ partial waves at the position  
$(2380\pm10)-i(40\pm5)$ MeV, which is in full agreement with the findings in the
two-pion production reactions \cite{prl2014,RWnew,npfull}. The bottom panel in
Fig.~\ref{fig-pn} displays the Argand diagram of the new solution AD14 for the
$^3D_3$ partial wave. It exhibits a pronounced looping of this partial wave in
agreement with a resonant behavior. The poles in $^3D_3$ and $^3G_3$
  partial waves have been reproduced in a theoretical study of
  nucleon-nucleon scattering within the constituent quark models of the
  Nanjing group \cite{WangNN}.  

Very recently also data for the differential cross sections of the $pn$
scattering in the region of $d^*(2380)$ could be extracted from the WASA data
base. It turned out that the new experimental data are perfectly described by
the AD14 partial-wave solution, which is a remarkable success putting further
confidence into the uniqueness and predictive power of this solution
\cite{pndsig}.

With this result the resonance structure observed in two-pion
production has been established as a genuine $s$-channel resonance in the
proton-neutron system. Due to its isoscalar character the notation $d^*(2380)$
has been chosen in analogy to the denotation for isoscalar excitations of the
nucleon.

\subsubsection{\it hadronic decay branchings of $d^*(2380)$}

From the various two-pion production measurements as well as from the $np$
scattering experiments and their analysis the branching ratios given in
Table 1 have been extracted for the hadronic decays of $d^*(2380)$
\cite{dibaryonreview,BR}. The decay into the not measured $nn\pi^+\pi^0$
channel has been taken to be identical to that into the $pp\pi^o\pi^-$ channel
by isospin symmetry. 

The observed $d^*$ decay
branchings into the diverse two-pion channels are consistent with
expectations from isospin decomposition \cite{BR} as well as explicit
QCD model calculations \cite{dong1}.

In a dedicated WASA search for the hypothetical decay $d^*(2380) \to
(NN\pi)_{I=0}$ no signal from $d^*(2380)$ could be sensed in the experimental
isoscalar single-pion production cross section, but an upper limit of 5$\%$ at
90$\%$ C.L. could be derived for such a branching
\cite{isoNNpi,isoNNpi_err}. Note that in Ref. \cite{isoNNpi} the upper limit was
given too high by a factor of two \cite{isoNNpi_err}. This result disfavors
strongly models predicting a molecular structure for $d^*(2380)$
\cite{Kukulin1,Kukulin2,GG1,GG2}, but is in full accord with a compact
hexaquark-$\Delta\Delta$ structure \cite{dongNNpi}. 

It should be pointed out that the successful reproduction of all hadronic
decay branchings of $d^*(2380)$ and its total width by the theoretical
calculations also means 
that all experimentally observed cross sections in the various hadronic
channels are understood theoretically in a quantitative manner.      
 
\begin{table}
\begin{center}
\begin{minipage}[t]{16.5 cm}
\caption{Branching ratios in percent of the $d^*(2380)$ decay into $pn$,
  $NN\pi$, $NN\pi\pi$ and $d\gamma$ channels. The experimental results
\cite{BR,isoNNpi,isoNNpi_err,Ishikawa1,Ishikawa2,Guenther} are compared to
results from a theoretical 
calculation 
  \cite{dong1,dongNNpi} starting from the theoretical $d^*$ wave function and
  including 
  isospin breaking effects. They are also compared to values expected from
  pure isospin recoupling of the various $NN\pi\pi$ channels. In the latter
  case the branching into the $d\pi^0\pi^0$ channel is normalized to the
  data.
}  
\label{tab:branching}
\end{minipage}
\begin{tabular}{llll}
\\ 
\hline\\

decay channel&experiment&theory \cite{dong1,dongNNpi}
&$NN\pi\pi$ isospin recoupling\\ 

\hline\\

$d\pi^0\pi^0$&~~$14\pm1$&~~~~~~~12.8&~~~~~~~~~~13\\
$d\pi^+\pi^-$&~~$23\pm2$&~~~~~~~23.4&~~~~~~~~~~26\\
$np\pi^0\pi^0$&~~$12\pm2$&~~~~~~~13.3&~~~~~~~~~~13\\
$np\pi^+\pi^-$&~~$30\pm5$&~~~~~~~28.6&~~~~~~~~~~32.5\\
$pp\pi^0\pi^-$&~~~$6\pm1$&~~~~~~~~4.9&~~~~~~~~~~~6.5\\
$nn\pi^+\pi^0$&~~~$6\pm1$&~~~~~~~~4.9&~~~~~~~~~~~6.5\\
$(NN\pi)_{I=0}$&~~~~~$<5~(90\%~C.L.)$&~~~~~~~~0.9&~~~~~~~~~~~--\\
$np$&~~$12\pm3$&~~~~~~~12.1&~~~~~~~~~~~--\\
$d\gamma$&~~$\approx$ 0.01&&\\
\hline\\
$\sum(total)$&~$103\pm7$&~~~~~~100&~~~~~~\\
\hline
 \end{tabular}\\
\end{center}
\end{table}

\section{Electromagnetic Excitation of $d^*(2380)$}

The electromagnetic decay channels are very interesting, since they offer the
possibility to excite the resonance also by photo- and electro-production,
respectively. The latter, in particular, offers the possibility to measure
that way the transition form-factor, which could give experimental access to
the size of $d^*(2380)$ and thus further clarify the question about the
structure of $d^*(2380)$. From the $\gamma$ decay of the $\Delta$ resonance
one may estimate that the cross sections for the processes $pn \to d^*(2380)
\to \Delta\Delta \to d\pi^0\gamma$ and $pn \to d^*(2380) \to \Delta\Delta \to
d\gamma\gamma$ are smaller than the hadronic decays by two and four orders of
magnitude, respectively. For the $d\gamma$ decay channel the situation is
presumably similar.

Possibly an indication of $d^*(2380)$ photo-excitation has been observed
already in the seventies in the photo-absorption reaction $\gamma d \to pn$ by
measuring the polarization of the ejected protons. After observation
\cite{Kamae1,Kamae2} of an anomalous 
structure in the proton polarization from deuteron photodisintegration Kamae
and Fujita \cite{Kamae} suggested the possible existence of a deeply bound
$\Delta\Delta$ system with $I(J^P) = 0(3^+)$ at $\sqrt s$ = 2.38 GeV with a
width of 160 MeV. Subsequent analyses based on an increased basis of
polarization measurements yielded the possible existence of at least two
dibaryon resonances with either $I(J^P) = 0(3^+)$ or $0(1^+)$ at $\sqrt
s$ = 2.38 GeV and $I(J^P) = 1(3^-)$ or $1(2^-)$ at $\sqrt s$ = 2.26 GeV with
widths of 200 MeV and above \cite{Ikeda1,Ikeda2}.

New measurements of the $\gamma
d \to pn$ reaction with polarized photons at MAMI --- measuring also for the
first time the polarization of the  emerging neutrons --- are consistent with
an excitation of $d^*(2380)$ in this process \cite{MB_Sigma,MB_pol}. At the 
photon energy corresponding to the $d^*(2380)$ mass both the previously
measured proton 
polarization \cite{Ikeda1,Ikeda2,JLAB} and the newly measured neutron
polarization peak at a scattering angle of 90 degrees in the center-of-mass
system with reaching a polarization of $P_y$=-1. This extreme value means that
the final pn system must be 
in a spin triplet state as required for a $d^*(2380) \to pn$ decay. The
measured \cite{MB_pol} energy dependence of $P_y$ at 90 degrees is in
agreement with a Lorentzian energy dependence having the width of $d^*(2380)$. 

Very recently also first data for the $\gamma d \to d\pi^0\pi^0$ reaction
appeared reaching in energy down to the $d^*(2380)$ region. The measurements
conducted at ELPH, Japan, only reach down to the high-energy side of the
$d^*(2380)$ region \cite{Ishikawa1,Ishikawa2}. Measurements performed at MAMI
reach even 
below the $d^*$ region \cite{Guenther}. Both measurements have to fight
heavily with background contributions at the lowest energies. However, both
measurements find a surplus of cross section in the $d^*$ region in comparison
to the state-of-the-art calculations of Fix and Arenhövel \cite{Fix,Egorov},
which describe the data very well above $\sqrt s$=2.40 GeV. A $d^*$
cross section of about 20 nbarn provides a good description of both data sets
in the $d^*$ region. This is just four orders of magnitude smaller than the
cross section in the corresponding hadronic channel providing an
electromagnetic branching of $10^{-4}$ for the $d^*(2380) \to d\gamma$
transition --- in agreement with the estimate given above. First attempts to
understand the photo-absorption process $\gamma d \to d^*(2380)$ theoretically
\cite{dongelm,MBelm} provide cross sections, which are still too low by an
order of magnitude.

\section{Status of Theoretical Work on $d^*(2380)$}

There have been a huge number of dibaryon predictions in the past. The oldest
one dates back to 1964, when based on SU(6) symmetry considerations Dyson and
Xuong \cite{Dyson} predicted six non-strange dibaryon states $D_{IJ}$, where the
indices denote isospin I and spin J of the particular state. They
identified the two lowest-lying states $D_{01}$ and $D_{10}$ with the deuteron
groundstate and the virtual $^1S_0$ state, respectively, the latter being
known from low-energy $NN$ scattering and final-state interaction. Identifying
in addition the third state $D_{12}$ with the at that time already debated
resonance-like structure at the $N\Delta$ threshold having $I(J^P)=1(2^+)$,
they fixed all parameters in their mass formula. As a result they predicted a
state $D_{03}$ with just the quantum numbers of $d^*(2380)$ at a mass of 2350
MeV, which is remarkably close to the now observed $d^*$ mass. The remaining
predicted states are $D_{21}$ and $D_{30}$. Due to their isospins I = 2 and 3
they are $NN$-decoupled. According to Dyson and Xuong they should have masses
very similar to those of $D_{12}$ and $D_{03}$ , respectively.
 
On the one hand it appears very remarkable, how well the prediction of Dyson
and Xuong works. On the other hand this may not be too surprising, because we
know since long that the mass formulas for baryons and mesons derived from
symmetry breaking considerations also do remarkably well, if a few
phenomenological parameters are adjusted to experimental results.

  Oka and Yazaki were the first to apply the nonrelativistic quark model to
  the problem of the nuclear force. They demonstrated that the 
  interplay between the Pauli principle and the spin-spin interaction between
  quarks leads to a strong short-range repulsion between nucleons, but to an
  attractive force for a $\Delta\Delta$ system with $I(J^P) = 0(3^+)$
  \cite{OkaYazaki}. 

Terry Goldman, Fan Wang and collaborators \cite{Goldman} pointed out later
that a $\Delta\Delta$ configuration with the particular quantum numbers
of $d^*(2380)$ must have an attractive interaction due to its special
  color-spin structure, so that any model 
based on confinement and effective one-gluon exchange must predict the
existence of such a state --- the "inevitable dibaryon" as they called it.
  In their quark-delocalization and color-screening model (QDCSM) they
  initially predicted a binding energy of 350 MeV relative to the nominal
  $\Delta\Delta$ threshold, but approached the experimental value in more
  recent work \cite{Ping,Wang,WangNN}.
 
In fact, many groups calculated meanwhile such a state at
actually similar mass based either on quark-gluon
\cite{Ping,Wang,Yuan,Li,LiShenZhang,Huang,Dong,Chen,Nijmegen,Mulders,MuldersThomas,Saito,An} 
or hadronic interactions \cite{GG1,GG2,Kamae}. 
  Already the early bag-model calculations of the Nijmegen theory group
  \cite{Nijmegen,Mulders} including the work of Mulders
  and Thomas \cite{MuldersThomas} and also Saito \cite{Saito} predicted
  $d^*(2380)$ at about the correct mass. However, in these calculations also
  numerous other unflavored dibaryon states were predicted, which have not been
  observed (at least so far) or which have been observed at a
  significantly different mass like, {\it e.g.}, the $D_{12}$ state. 

Another correct real prediction, {\it i.e.} a prediction before the
experimental observation of $d^*(2380)$, is the one by the IHEP theory group
led by Z. Y. Zhang, who studied this state in the chiral SU(3) quark model
within the resonating group method \cite{Yuan}. This work and follow-up
  investigations of this group \cite{Li,LiShenZhang,Huang,Dong} include
  the concept of "hidden color". Hidden-color
six-quark states are a rigorous first-principle prediction of SU(3) color
gauge theory. Six quark color-triplets $3_c$  combine to five different
color-singlets in QCD and will significantly decay to $\Delta\Delta$ as shown
in Refs. \cite{Brodsky,Brodsky1}. Problems related to the application of
hidden color in multi-quark systems have been discussed by Fan Wang {\it et
  al.} \cite{Wanghidden}. They point out, while the $\Delta\Delta$ and
hidden-color configurations are orthogonal for large separations between the
two quark clusters, they loose their orthogonality, when they start to 
overlap at small separations.
Another point of caution has been noted by F. Huang and W.L. Wang in Ref. 
\cite{HuangNN}. In this work they study the masses of octet and decuplet
baryon ground states, the deuteron binding energy as well as the $NN$
scattering phase shifts (for $J\leq$ 6) below the pion-production threshold
within a chiral SU(3) quark model. They demonstrate that all these can be well
described, if the consistency requirement 
for the single-baryon wave functions to satisfy the minima of the Hamiltonian
are strictly imposed in the determination of the model parameters. In earlier
quark-model calculations usually the nucleon is set to be at the minimum of
the Hamiltonian by a particular choice of the model parameters. In consequence,
the $\Delta$, which is of different size, is not stable against its size
parameter in the wave function, {\it i.e.} its wavefunction is not the
real solution of the Hamiltonian. Hence one needs to be very careful when
extending the model from the study of the NN interaction to other
baryon-baryon systems and one may need to introduce additional channels like
the hidden-color channel to lower the energy of the $\Delta\Delta$
system. These channels might not be the physical ones, but are partially
needed to change the internal wave function of the single
$\Delta$. Therefore one has to be cautious in explaining the configuration
structure of the coupled $\Delta\Delta$-hidden-color system. Hence the IHEP
result of 2/3 hidden-color components in $d^*(2380)$ has to be taken with some 
caution with regard to its interpretation of the configuration of
$d^*(2380)$. An improved calculation for $d^*(2380)$ with a consistent
treatment of the $\Delta\Delta$ system is in progress \cite{Huangpriv}.

Recently also a diquark model has been proposed for $d^*(2380)$
\cite{Shi}. In this work it is assumed that $d^*(2380)$ is composed of
three vector diquarks and its mass is calculated by use of an effective
Hamiltonian approach. Surprisingly, in this rough and simple model both mass
and width (see next subsection) come out in good agreement with the experimental
data.  In a subsequent paper \cite{GalKarliner} Gal and Karliner questioned
the applicability of diquark models in the light-quark sector by demonstrating 
that the use of the effective Hamiltonian with parameters given in
Ref. \cite{Shi} leads to masses for deuteron- and virtual-like states,
which are 200-250 MeV above the physical deuteron and the virtual $^1S_0$
state. However, as pointed out in a reply, the latter two states interpreted
as three-axial-vector-diquark states reside in spin-flavor multiplets
different from the one of $d^*(2380)$ and need a Hamiltonian with more
interactions included \cite{Huangpriv}. 

As a historic side remark we note that a diquark model for a deuteron-like
object had been proposed already some time ago \cite{demon}, where three
scalar diquarks were coupled in relative $P$-wave to an isoscalar $J^P = 0^-$
object, the so-called "demon deuteron" possessing a highly suppressed
decay. In this context the data for $np \to d\pi^+\pi^-$, which were available
at that time and which 
indicated a peak in the total cross section around $\sqrt s \approx$ 2.3 GeV
and exhibited the ABC effect (see section 2.3.1), were interpreted as evidence
for the existence of such a "demon deuteron". As we know now, this turns out
to be just the place, where $d^*(2380)$ was found instead.

Meanwhile also a QCD sum rule study has found this state at the right mass
\cite{An}, whereas another QCD-based work without any inclusion of hadron
degrees of freedom can construct such a state as a compact object only at much
higher masses \cite{Park}. 

Most recently first lattice QCD calculations for $d^*(2380)$ were presented by
the HAL QCD collaboration \cite{halqcd,halqcd_new} finding evidence for a bound
$\Delta\Delta$ system with the quantum numbers of $d^*(2380)$. In these
calculations the pion mass is still unrealistically large, because
$\Delta(1232)$ has to be assumed to be a stable particle, in order to
make such calculations feasible at present. Therefore the lattice
results 
were recently extra\-polated down to the real pion mass by methods based on
Effective Field Theory with the result that indeed such a bound state is
likely to exist \cite{Haidenbauer}. 

Gal and Garcilazo also obtained this
state at the proper mass in recent relativistic Faddeev calculations
based on hadronic interactions  within a baryon-baryon-pion system
\cite{GG1,GG2} and assuming a decay $d^*(2380) \to D_{12}\pi \to \Delta N
\pi$. Such a decay was also investigated by Kukulin and Platonova 
\cite{Kukulin1,Kukulin2}. 

\subsection{\it the width issue}

More demanding than the mass value appears to be the reproduction of the small
decay width of $d^*(2380)$. As worked out in a paper together with Stanley
Brodsky \cite{BBC}, the small width points to an unconventional origin, possibly
indicating a genuine six-quark nature. With the dominant decay being
$d^*(2380) \to \Delta\Delta$ one would naively expect a reduction of the decay
width from $\Gamma_{\Delta\Delta}$=240 MeV to 160 MeV for a $\Delta\Delta$
system bound by 90 MeV – using the known momentum dependence of the width of
the $\Delta$ resonance. This is twice the observed width. On the other hand,
if $d^*(2380)$ is a genuine six-quark state, we need to understand its large
coupling to $\Delta\Delta$. This can be explained, if one assumes that
$d^*(2380)$ is dominated by "hidden-color" configurations.

So far there have been five predictions for the decay width based 
on Faddeev calculations \cite{GG1,GG2}, quark-model calculations
\cite{dong1,Wang,Huang,Dong,Ping,Huang1,Lue,Shi} or some general
considerations \cite{Niskanen}. A width of 160 MeV as
discussed above is also obtained initially in the quark-model calculations of
Fan Wang et al. \cite{Ping}. By accounting in addition for correlations in a
more detailed treatment they finally arrive at 110 MeV \cite{Wang}. A
similar width is obtained in the Faddeev calculations \cite{GG1,GG2}. For a
resonance mass of 2383 MeV they obtain a width of 94 MeV for the decay into
all experimentally observed $NN\pi\pi$ decay channels. Adding the decay width
into the $pn$ channel, which they cannot account for in their model, leads
finally to a width of 104 MeV.  

The quark-model calculations of the IHEP group, which include
hidden-color configurations, as discussed in  Refs. \cite{Brodsky,Brodsky1,BBC},
arrive at the experimentally observed width
\cite{dong1,Huang,Dong,Huang1,Lue,Donghexa}. In these calculations the $d^*(2380)$
hexaquark of size 0.8 fm contains about 67$\%$ hidden-color components, which
cannot decay easily and hence reduce the width to the experimental value.
 
Also the diquark model of Shi, Huang and Wang \cite{Shi} reproduces the
observed narrow width. Here the width is naturally explained by the large
tunneling suppression of a quark between a pair of diquarks. Again, Gal and
Karliner \cite{GalKarliner} question this result arguing that in the
calculation of the decay an isospin-spin recoupling factor of 1/9 has been
overlooked, which would reduce the width to less than 10 MeV. However, in a
reply Shi and Huang point out that such a recoupling factor appears only in
uncorrelated quark models, but not in the diquark model \cite{Huangpriv}.

For completeness we mention here also the recent work of Niskanen
\cite{Niskanen}, who considers the energy balance in $\Delta N$ and
$\Delta\Delta$ systems. He arrives at the surprising conclusion that both
these systems should have widths, which are substantially smaller than the
width of the free $\Delta$ at the corresponding mass. This conclusion is not
only counterintuitive as he also notes, but also in sharp contrast to the
experimental results. {\it E.g.}, for $d^*(2380)$ he obtains a width of about
40 MeV and for $D_{12}$ a width of about 75 MeV, {\it i.e.} in both cases much
smaller than observed experimentally. Such Fermi motion considerations have
been taken up recently also by Gal \cite{Gal} for the discussion of the
expected size of the 
$\Delta\Delta$ configuration of $d^*(2380)$. In Ref. \cite{Meson2018} it is
demonstrated that such considerations lead to conflicts with the observed mass
distributions. What is observed in these spectra are $\Delta$s of mass 1190
MeV with a width of 80 MeV --- as expected from the mass-width relation of a
free $\Delta$. This is in line with the expectation that during the decay
process $d^*(2380) \to \Delta\Delta$ the distance between the two $\Delta$
increases continuously eliminating thus the Fermi motion finally and putting
mass and width of the $\Delta$s back to their asymptotic values.

\subsection{\it hexaquark versus molecular structure}

If the scenario of the models, which correctly reproduce the
experimental width, is true, then the unusually small decay width of
$d^*(2380)$ 
signals indeed an exotic character of this state and points to a compact
hexaquark nature of this object as discussed by the IHEP group
\cite{Huang,Huang1,Lue,Donghexa}. In fact, the IHEP calculations as well as
the quark-model calculations of the Nanjing group \cite{Wanghidden} give a value
as small as 0.8 - 0.9 fm for the root-mean-square radius of $d^*(2380)$, {\it
  i.e.}, as small as the nucleon. Also latest 
lattice QCD calculations provide values in the same range
\cite{halqcd}. Actually, such values appear to be not unreasonable, if one
uses just the uncertainty-relation formula \cite{Krane} 

\begin{equation}
R \approx \hbar c / \sqrt{2\mu_{\Delta\Delta} B_{\Delta\Delta}} \approx 0.5 fm
\end{equation}

for an order-of-magnitude estimate of the size of a $\Delta\Delta$ system bound
by $B_{\Delta\Delta}$ = 80 MeV and a reduced mass $\mu_{\Delta\Delta} =
m_{d^*(2380)}$.  

In contrast, the Faddeev calculations of Gal and Garcilazo give a
molecular-like $D_{12}-\pi$ structure with radius of about 2 fm
\cite{Gal,dong2}, {\it i.e.} as large as the deuteron. 
 Unfortunately, as
 demonstrated recently \cite{dibaryonreview,Gal}, the $d^*$ decay branchings
 into the various $NN\pi\pi$ channels based on isospin coupling turn out to be
 identical for the routes $d^* \to \Delta\Delta \to NN\pi\pi$ and $d^* \to
 D_{12}\pi \to NN\pi\pi$, respectively, and hence do not discriminate between
 these two scenarios. Fortunately, however, there is a way out by looking at a
possible decay into the single-pion channel. In leading order such a decay is
forbidden for $d^* \to \Delta\Delta \to (NN\pi)_{I=0}$ . The consideration of
higher order terms gives a branching of less than 1$\%$ \cite{dongNNpi}. The
situation is much different for an intermediate $D_{12}\pi$ configuration,
since $D_{12} \to NN$ has a branching of 16 - 18$\%$ \cite{RWnew,Arndt}. Hence
the $d^* \to NN\pi$ decay ought to have the same branching in this 
scenario. However, exactly this has been excluded by the dedicated WASA single-pion
production experiment \cite{isoNNpi,isoNNpi_err}. In consequence of this
experimental 
result Avraham Gal proposed a mixed scenario, where the main component of
$d^*(2380)$ consists of a compact core surrounded by a dilute cloud of
$D_{12}-\pi$ structure \cite{Gal}. 

Summarizing, in the present discussion about the nature of $d^*(2380)$ it is no
longer the question about its existence itself, but about its structure. Is it
a dilute molecular-like object or is it a compact hexaquark object? The
measured decay properties of $d^*(2380)$ clearly favor the latter.

\section{ Recent Searches for Dibaryonic States in the $\Delta(1232)N$, $N^*(1440)N$
and $\Delta(1232)\Delta(1232)$ Regions.}

\subsection{\it $\Delta N$ threshold region}

Stimulated by the success in establishing $d^*(2380)$ as the first narrow
dibaryon resonance of non-trivial nature, new experiments have been
started recently to search for other possible dibaryon resonances.
With the ANKE detector at COSY the $pp \to pp\pi^0$ reaction was studied
with polarized protons over a large energy range $\sqrt s$ = 2040 - 2360
MeV and under the kinematical condition that the emitted proton pair
is in the relative $^1S_0$ state \cite{ANKE}. Thus these measurements are
complementary to the ones of the $pp \to d\pi^+$ reaction, where the
nucleons bound in the deuteron are in relative $^3S_1$ state --- aside
from the small $D$-wave admixture in the deuteron.

In the partial wave analysis of their data the ANKE collaboration
finds the $^3P_0 \to $$^1S_0s$ and $^3P_2 \to$$^1S_0d$ transitions to be
resonant 
peaking at 2201(5) and 2197(8) MeV, respectively, with widths of 91(12)
and 130(21) MeV, respectively. The resonance parameters point to $\Delta N$
threshold states with $I(J^P) = 1(0^-)$ and $1(2^-)$, respectively, where the
two constituents $N$ and $\Delta$ are in relative $P$ wave. The particular
signature of the $^3P_0 \to$$^1S_0s$ and $^3P_2 \to$$^1S_0d$ transitions is that
they constitute proton spinflip transitions, which cause concave shaped
pion angular distribution in contrast to the conventional convex
shaped ones. This peculiar behavior was noted already before in
PROMICE/WASA \cite{Jozef} and COSY-TOF \cite{evd} measurements of the $pp \to
pp\pi^0$ 
reaction at energies near the pion production threshold providing thus
first hints for a resonant behavior of these partial waves. The masses
of these $p$-wave resonances are slightly above the nominal $\Delta N$ mass.
This is understood to be due to the additional orbital motion \cite{ANKE}.

For the $I(J^P) = 1(2^-)$ resonance corresponding to the $^3P_2$ $NN$-partial
wave a pole had been found already before
in SAID partial-wave analyses of data on $pp$ elastic scattering and the
$pp \rightleftharpoons d\pi^+$ reaction \cite{FA91,Arndt}. In these analyses
also evidence for 
poles in $^3F_3$ and $^3F_4-$$^3H_4$ partial waves have been found near the
$\Delta N$ 
threshold, though these evidences appear much less pronounced than for
the above cases. These poles would correspond to states with $I(J^P) =
1(3^-)$ and $1(4^-)$.

Kukulin and Platonova have demonstrated recently that by accounting
for the isovector $P$-wave resonances as well as the dominant isovector $2^+$
resonance the $pp \rightleftharpoons d\pi^+$ cross section and polarization
observables can be described quantitatively for the first time with form-factor
cut-off parameters, which are consistent to those obtained in elastic scattering
descriptions \cite{Kukulin3P2}. 

Not coupled to the $NN$ channel, but in the region of the $\Delta N$
threshold, there is supposed to be another state with quantum numbers mirrored
to those of the $I(J^P) = 1(2^+)$ state. This state with $I(J^P) ) = 2(1^+)$
--- first predicted by Dyson and Xuong in 1964 \cite{Dyson} and denoted by
$D_{21}$ 
-- is decoupled from the elastic $NN$ channel due to its isospin I = 2.
Hence, it only can be produced in $NN$-initiated reactions associatedly,
e.g., by the $pp \to D_{21}\pi^- \to pp\pi^+\pi^-$ reaction. Though the total
cross section of this two-pion production channel runs smoothly over the
$\Delta N$ threshold region, it was noted recently that its slope there is
not in accord with isospin relations between this channel and the $pp
\to pp\pi^0\pi^0$ channel. Whereas the latter cannot contain a $D_{21}$ resonance
excitation due to Bose symmetry, the first can do so. And indeed, a
detailed analysis of WASA-at-COSY $pp \to pp\pi^+\pi^-$ data revealed
pronounced differences in invariant mass spectra and angular
distributions associated with $\pi^+$ or $\pi^-$. These cannot be understood
by conventional $t$-channel reaction mechanism, however, quantitatively
described by the presence of $D_{21}$ with m = 2140(10) MeV and $\Gamma$ =
110(10) MeV \cite{D21,D21long}. Aside from the prediction of Dyson and Xuong
also Gal and Garcilazo obtain this state with about the same mass and width
\cite{GG2}, whereas the Nanjing group does not get enough binding in their
calculation for the formation of a bound state \cite{NanjingDeltaN}.

\subsection{\it  N*(1440)N region}

In contrast to $\Delta N$ resonances, which can couple solely to isovector
$NN$ channels, $N^*N$ resonances can connect both to isoscalar and
isovector $NN$ channels. So the most likely configurations, where $N^*$ and
$N$ are in relative $S$-waves, can couple to $^1S_0$ and $^3S_1$ $NN$-partial
waves possessing the quantum numbers $I(J^P) = 1(0^+)$ and $0(1^+)$,
respectively. 

In fact, evidence for the existence of such states has been found just
recently in $NN$-initiated single- and double-pion production. In a
study dedicated initially to the search for a decay $d^*(2380) \to NN\pi$
--- see section 2.3.4 --- the isoscalar part of the single-pion
production was measured in the $d^*$ resonance region covering
also the $N^*(1440)N$ excitation region \cite{isoNNpi}. As a result, the
isoscalar total cross section is observed to increase monotonically from the
$NN\pi$ threshold up to $\sqrt s \approx$ 2.32 GeV --- as also expected for a
conventional $N^*$ excitation process mediated by $t$-channel meson
exchange. However, one would expect in such a case that the cross
section keeps rising as the beam energy is further increased. Instead,
the measurements beyond 2.32 GeV exhibit a decreasing cross
section forming thus a bell-shaped energy excitation function
for the isoscalar total cross section. Since the simultaneously
measured isoscalar $N\pi$-invariant-mass distribution is in accord with
an excitation of the Roper resonance $N^*(1440)$ \cite{isoNNpi}, the
observation has to be interpreted as evidence for a $N^*N$ resonance
\cite{NstarN,NstarN_exp}. 
We deal here with a state below the nominal $N^*N$ mass of $m_{N^*} + m_N$ =
2.38 GeV. Therefore  $N^*$ and $N$ have to be in relative $S$-wave and the
quantum numbers of this resonance must be $I(J^P) = 0(1^+)$. {\it I.e.}, it is
fed by the $^3S_1$ partial wave in the $NN$-system. 

Since the Roper resonance decays also by emission of two pions, this
$N^*N$ structure could possibly be seen in isoscalar two-pion production,
too. This is particularly true for the $pn \to d\pi^0\pi^0$ reaction, where the background
of conventional processes is lowest and where also $d^*(2380)$ is
observed best. As seen in Figs. 1 and 2 there is, indeed, a small surplus of
cross section in the region of $\sqrt s \approx$ 2.3 GeV (black filled circles in
Fig.~2), i.e., at the low-energy tail of the $d^*(2380)$ resonance, which
could be related to the isoscalar $N^*N$ state.

Isospin decomposition of data on various $pp$-induced two-pion
production channels had revealed already some time ago that the Roper
$N^*(1440)$ excitation process exhibits a bump-like structure there, too,
peaking in the region of the $N^*N$ mass \cite{iso}. Since the initial
$pp$-system is of isovector character, the observed structure must correspond
to a $N^*N$ state with $I(J^P) = 1(0^+)$ formed by the $^1S_0$
partial wave in the initial $pp$ channel.

Both resonance structures peak around 2320 MeV and have a width of
$\Gamma \approx$ 150 MeV. These values conform with the pole parameters of 1370
- i 88 MeV for the Roper resonance, however, not with its Breit-Wigner
values of $m \approx$ 1440 MeV and $\Gamma \approx$ 350 MeV \cite{PDG}. If
the latter mass value is taken for the nominal $N^*N$ mass, then the two
$N^*N$ structures appear to be bound by about 70 MeV, which could explain that
the observed width is smaller than typical for a Roper excitation. In fact,
the formation of a $N^*N$ resonance state would also explain the observation
that in nucleon-accompanied Roper excitations like, e.g., in hadronic $J/\Psi
\to \bar{N}N\pi$ decay \cite{BES} and $\alpha N$ scattering
\cite{Morsch1,Morsch2}, this excitation is always seen with values close to
its pole parameters, but not as expected with its Breit-Wigner values.

\subsection{\it $\Delta\Delta$ region}

In addition to the peak for the excitation of the $d^*(2380)$ resonance
at $\sqrt s$ = 2.37 GeV there appear two further peaks at 2.47 and 2.63
GeV in the total cross section of the $\gamma d \to d\pi^0\pi^0$ reaction, as
measured recently both at ELPH \cite{Ishikawa2} and at MAMI
\cite{Guenther}. Whereas 
conventionally these two bumps are explained to belong to
electromagnetic excitations of the nucleon in the so-called second and
third resonance region \cite{Fix,Egorov}, the collaboration at ELPH
demonstrates that 
the measured angular distributions are not in accord with such a
quasifree reaction process, but rather with a process for the
formation of isoscalar dibaryon resonances with masses m = 2469(2) and
2632(3) MeV and widths of $\Gamma$ = 120(3) and 132(5) MeV, respectively
\cite{Ishikawa2}. No spin-parity assignments are given, but the
$d\pi$-invariant mass spectra suggest a decay of these putative resonances via
$D_{12}$, the isovector $2^+$ state near the $\Delta N$ threshold. Whereas the
peak at 2.63 GeV is beyond the 
energy range measured at WASA in the $pn \to d\pi^0\pi^0$ reaction, the peak
at 2.47 GeV is still within this range. Since the peak cross
section at 2.47 GeV is roughly double as high as that for $d^*(2380)$ at
2.37 GeV, one would naively expect a similar situation also in the
hadronic excitation process measured by WASA. But nothing spectacular
is seen around 2.5 GeV in the WASA measurements. The observed small
cross section in this region --- see Fig. 2 --- is well understood by
the conventional $t$-channel $\Delta\Delta$ process as indicated in Fig. 1.
A possible way out could be the conception that similar to the
situation with the Roper resonance also the higher-lying nucleon
excitations undergo a kind of molecular binding with the neighboring
nucleon at their threshold. Since in $NN$-induced reactions the
excitation of the hit nucleon into states of the second and third
resonance region has a much smaller cross section \cite{Teis} than the
conventional $\Delta\Delta$ process, it could be understandable, why WASA
does not observe the peak at 2.47 GeV seen in $\gamma$-induced $\pi^0\pi^0$
production, where the $\Delta\Delta$ process is not possible.

Five out of six dibaryonic states predicted by Dyson and Xuong \cite{Dyson} in
1964 have been found meanwhile with masses even close to the predicted
ones --- if the interpretation of the WASA data as evidence for $D_{21}$
  is correct, see section 5.3. Therefore it appears very intriguing to
investigate, whether 
also the sixth one exists, perhaps also close to its predicted mass
value. This $NN$-decoupled state $D_{30}$ with $I(J^P)=3(0^+)$, i.e. with quantum
numbers mirrored to those of $D_{03}= d^*(2380)$ and of $\Delta\Delta$ nature,
too, is particularly difficult to find, since one needs at least two
associatedly produced pions, in order to produce it in $NN$-initiated
reactions.

An attempt to search for it in WASA data for the $pp \to pp\pi^+\pi^+\pi^-\pi^-$
reaction was undertaken recently \cite{I3dibaryon}. No stringent signal of
such a 
state was observed in these data and only upper limits for its
production cross section could be derived, because the theoretical
description of conventional processes for four-pion production is not
well known so far. However, it could be shown that the upper limit is
at maximum for the combination m = 2.38 GeV and $\Gamma$ = 100 MeV. {\it
  I.e.}, if this state really exists, then this mass-width combination is most
likely.

Since this mass is compatible with the $d^*(2380)$ mass, it would agree
perfectly with the prediction of Dyson and Xuong, who get equal masses
for both these states. Also other theoretical studies \cite{GG2,Li,LiShenZhang} find $D_{30}$
to lie in this mass region.

\section{ Remarks on Flavored Dibaryons}

Despite of numerous experimental attempts no single dibaryon candidate
could be established firmly so far in the flavored quark sector. Most
of experiments were carried out in the strange sector, in particular
searching for the so-called $H$-dibaryon, a bound $\Lambda\Lambda$ state
predicted 1977 by Jaffe \cite{Jaffe}. For a recent review see, {\it e.g.}
Ref.~\cite{dibaryonreview}. According to very recent lattice QCD
  simulations close to the physical point ($m_\pi$ = 146 MeV, $m_K$ = 525 MeV)
  performed by the HAL QCD collaboration, there is no bound or resonant 
  $H$-dibaryon around the $\Lambda\Lambda$ threshold \cite{halqcdH}. However,
  a possible $H$ resonance close to the $\Xi N$ threshold cannot yet be 
  excluded by these calculations.
 
  The dibaryon search in the strange sector has received a new push, after the
lattice QCD calculations by the HAL QCD collaboration keep predicting 
slightly bound $\Omega\Omega$  and $\Omega^- p$ systems
\cite{halqcdOmegaOmega,halqcdOmegaN}. The latter result is also in accord with
quark model calculations by the Nanjing group
\cite{NanjingOmegaN}. First measurements of the $\Omega^- p$ correlation
function by the STAR experiment at RHIC give first hints that the scattering
length is indeed positive in favor of a bound state in this system \cite{STAR}. 

An established unusual structure found in the strange sector is a
narrow spike at the $\Sigma N$ threshold, conventionally interpreted as a
cusp effect \cite{dibaryonreview,KE,SJ,Muenzer}. But also a possible $\Lambda
N$ state has been
discussed, for a recent review on this subject, see, {\it e.g.},
Ref.~\cite{Machner}. 

At JPARC experiments are going on to search for a bound $ppK^-$ system.
Recent results are in favor of the existence of such a system \cite{JPARCE15},
however, a definite confirmation has to be still awaited.

  Lately particular attention was paid to the charm and beauty sector,
  where tetra- and pentaquark system were observed recently. This finding
  suggests that in these sectors the attraction is again large enough to form
  dibaryons. Such expectation has been supported meanwhile by numerous model
  calculations of increasing sophistication. {\it E.g.}, Fr\"omel {\it et al.}
  \cite{Froemel} started out with well-established phenomenological
  nucleon-nucleon potentials applying quark-model scaling factors for scaling
  the strengths of the different interaction components and obtained first
  indications of deuteron-like bound states between nucleons and singly- as
  well as doubly-charmed hyperons. On the other hand a quark-model
  investigation of doubly-heavy dibaryons does not find any bound or
  metastable state there \cite{Valcarce1,Valcarce2}. Another quark-model study
  finds four sharp resonance states near the $\Sigma_c N$ and $\Sigma_c^* N$
  thresholds \cite{OkaYcN}. A recent lattice QCD 
  study based on the HAL QCD method \cite{halqcdLcN} comes to the conclusion
  that the attraction in the $\Lambda_c N$ system is not strong enough to form
  a bound system.

  Within the
  one-boson-exchange model Zhu {\it et al.} have undertaken a systematic
  investigation of single- and doubly-heavy baryon-baryon combinations
  \cite{Zhu1,Zhu2,Zhu3,Zhu4,Zhu5,Zhu6}. They find several candidates of
  loosely bound molecular states in the $\Xi_{cc} N$ system \cite{Zhu4}, for
  loosely bound deuteron-like states in the $\Xi_c\Xi_c$ and $\Xi_c^{'}\Xi_c^{'}$
  \cite{Zhu1} as well as in $\Lambda_c\Lambda_c$ and $\Lambda_b\Lambda_b$
  \cite{Zhu2} systems. They also get binding solutions for the
  $\Xi_{cc}\Xi_{cc}$ system \cite{Zhu3} and a pair of spin-3/2 singly-charmed
  baryons with the striking result that molecular states of
  $\Omega_c^*\Omega_c^*$ might be even stable \cite{Zhu5}. Possible molecular
  states composed of doubly charmed baryons are also investigated and good
  candidates have been found. But it is also demonstrated that coupled-channel
  effects can be important for the question, whether there is a binding
  solution or not \cite{Zhu6}.

  Unfortunately there are no experimental results yet. But with such many
  promising candidates it will be very interesting to watch future experiments
  in this area of charm and beauty.

\section{ Summary}
 
\begin{table}
\begin{center}
\begin{minipage}[t]{16.5 cm}
\caption{Unflavored Dibaryon Resonances (including candidates discussed in
  this review) below or near $\Delta(1232)N$, $N^*(1440)N$ and
  $\Delta(1232)\Delta(1232)$ thresholds. The resonance 
  states are characterized by their isospin $I$, spin-parity $J^P$, involved
  $NN$ partial wave $(^{2S+1}L_J$ with spin $S$, orbital angular momentum $L$
  and total angular momentum $J$), mass $m$ and width $\Gamma$. The column
  "evidence" shows a star rating based on the collected experimental evidence
  for the envisaged state (or candidate). 
The column "structure" denotes the asymptotic configuration of the particular
state in the course of its decay into the hadronic channels --- or also its
hindrance in case of hidden color. The column
"experimental information" summarizes recent references to corresponding
experimental work. The column "theoretical calculation" gives references to
theoretical work for the particular dibaryon state.
}  
\label{tab:states}
\end{minipage}
\begin{tabular}{lllllllll}
\\ 
\hline\\

I&$J^P$&$(^{2S+1}L_J)_{NN}$
&m (MeV)&$\Gamma$ (MeV)&evidence&asympt.&experimental&theoretical \\ 
&&&&&&structure&information&calculation \\

\hline\\
0&$1^+$&$^3S_1-$$^3D_1$&2320(10)&150(30)&***~~&$N^*N$&\cite{NstarN,NstarN_exp}&\cite{NstarN}\\
0&$3^+$&$^3D_3-$$^3G_3$&2370(10)&~70(5)&*****&$\Delta\Delta
$&\cite{dibaryonreview,prl2009,prl2011,isofus,prl2014}&\cite{Dyson,Goldman,Wang,Li}\\
&&&&&&$\rightleftharpoons$&\cite{BR,pp0-,np00,np+-}&\cite{Huang,dong1,dongNNpi,dongelm}\\
&&&&&&$hexaquark$&\cite{npfull,pndsig,isoNNpi,isoNNpi_err}&\cite{Yuan,Chen,Mulders,Saito}\\
&&&&&&$hidden~color$&\cite{Ishikawa1,Guenther,MB_Sigma,MB_pol}&\cite{An,Park,Shi,GalKarliner}\\
&&&&&&&&\cite{Kukulin1,Kukulin2,Kamae,GG1}\\
&&&&&&&&\cite{GG2,Gal,halqcd_new,Haidenbauer}\\
&&&&&&&&\cite{BBC,Kukulin4}\\
0&?&?&2469(2)&120(3)&*&?&\cite{Ishikawa2}& \\
0&?&?&2632(3)&132(5)&*&?&\cite{Ishikawa2}&  \\
\\
1&$0^+$&$^1S_0$&2315(10)&150(20)&*&$N^*N$&\cite{NstarN}&\cite{NstarN} \\
1&$0^-$&$^3P_0$&2201(5)&~91(12)&***&$\Delta N$&\cite{ANKE}& \\
1&$2^-$&$^3P_2-$$^3F_2$&2197(8)&130(21)&****&$\Delta N$&\cite{FA91,Arndt,ANKE}&\cite{Kukulin3}\\
1&$2^+$&$^1D_2$&2146(4)&118(8)&****&$\Delta
                                     N$&\cite{dibaryonreview,FA91,hos1,hos2}&\cite{Dyson,GG2,dong2,NanjingDeltaN}\\
  &&&&&&&&\cite{Kukulin3}\\
1&$3^-$&$^3F_3$&2183(?)&158(?)&**&$\Delta N$&\cite{FA91,Arndt}&\cite{Kukulin3} \\
1&$4^-$&$^3F_4-$$^3H_4$&2210(?)&156(?)&*&$\Delta N$&\cite{FA91,Arndt}& \\
\\
2&$1^+$&$^3P_1 + \pi$&2140(10)&110(10)&***&$\Delta N$&\cite{D21,D21long}&\cite{Dyson,GG2,NanjingDeltaN} \\
\\
3&$0^+$&$^1S_0 + \pi\pi$&2380??&100??&???&$\Delta\Delta
$&\cite{I3dibaryon}&\cite{Dyson,GG2,Wang,Li}\\
&&&&&&&&\cite{LiShenZhang}\\\\
\hline\\
 \end{tabular}\\
\end{center}
\end{table}


Following a trace laid by the ABC effect the WASA measurements of the
two-pion production in $np$ collisions revealed the existence of a narrow
dibaryon resonance, first observed in the $pn \to d\pi^0\pi^0$ reaction. As
it turned out, this has been the golden channel for the dibaryon discovery,
since the background due to conventional processes is smallest in this
channel. There was also no chance to discover it by previous experiments
in this channel, since there existed no other installation, which would
have been able to take reliable data for this reaction channel at the
energies of interest.

By subsequent WASA measurements of all possible hadronic decay channels this
dibaryon state could be established as a genuine $s$-channel resonance with
$I(J^P)=0(3^+)$ at 2370 - 2380 MeV. Its dynamic decay 
properties point to an asymptotic $\Delta\Delta$ configuration, bound by 80 -
90 MeV. Its width of only 70 MeV -- being more than three times smaller than
expected for a conventional $\Delta\Delta$ system excited by $t$-channel meson
exchange -- points to an exotic origin like hidden-color effects in the
compact hexaquark system.

Though the observation of such a state came for many as a surprise, it was
predicted properly already as early as 1964 by Dyson and Xuong and more
recently by Z. Y. Zhang {\it et al.}, who also can reproduce all measured
decay properties. Most recently this state has been also seen in  lattice QCD
calculations. 



Though there is meanwhile added evidence for a number of dibaryonic
states near the 
$\Delta N$ threshold --- all of them with large widths --- $d^*(2380)$ remains
so far the only established resonance with a surprisingly small width pointing
to a compact hexaquark structure of this state.

Key informations about the (unflavored) dibaryon states discussed in this
review are summarized in Table~2. For a number of them their existence is not
(yet) certain. 
The column "evidence" gives a star rating for the presently
collected experimental evidence of the envisaged state. It is based on the
authors' personal judgment and may serve just as a kind of guideline. The
experimentally best established one is certainly the isoscalar resonance
$d^*(2380)$ followed by the isovector $\Delta N$ near-threshold state with
$J^P = 2^+$.

The column "structure" denotes the asymptotic configuration of the particular
state in the course of its decay into the hadronic channels --- or also its
hindrance in case of hidden color. The column
"experimental information" summarizes recent references to corresponding
experimental work. The column "theoretical calculation" gives references to
theoretical calculations for the particular dibaryon state, be it predictions
or "postdictions".

\section{ Outlook}

From the measurements of the double-pionic fusion to $^3He$ \cite{3he} and
$^4He$ \cite{4he} we know that $d^*(2380)$ obviously survives in nuclear
surroundings. 
There are several other remarkable enhancements induced by $np$ pairs inside
nuclei. One is seen in di-electron pairs \cite{dls,hades} in heavy ion
collisions. This may be partly due to $d^*(2380)$ production inside nuclei
\cite{DLS}. Another is that the $np$ short-range correlation is found to be
about 20 times higher than the $pp$ in $(p,p’)$ and $(e,e’)$ scattering off
nuclei \cite{Piasetzky,Subedi,CLAS}. The question arises, whether an
intermediate formation of isoscalar dibaryon states like $d^*(2380)$ in the
course of the interaction between the nucleons in the nucleus may be an
explanation for this phenomenon. This would be in line with the
$NN$-interaction ansatz by Kukulin {\it et.al.}, where the short- and
intermediate range part of the $NN$-interaction is assumed to be due to
virtual $s$-channel dibaryon formation \cite{KukulinAnn,Kukulin3}. In fact, inclusion of
$d^*(2380)$ leads to a quantitative description of the $^3D_3 - $$^3G_3$ phase
shifts in both its real and imaginary parts \cite{Kukulin4}. Similar good
results are obtained for most of the partial waves with low orbital momentum,
when the $NN$-coupled dibaryon states given in Table~2 are included
\cite{NstarN,Kukulin3,Kukulin4}. 

Since $d^*(2380)$ appears to exist in nuclear matter, it can influence the
nuclear equation of state, especially in compact stellar objects like neutron
stars. A study finds $d^*(2380)$ to appear at densities three times the
saturation density and to constitute around 20$\%$ of the matter in the center
of neutron stars \cite{MB_nstar} depending on the assumed interaction of
$d^*(2380)$ with its surroundings \cite{MB_nstar1}.

Also, since dibaryons are bosons, one may think about a Bose-Einstein
condensate formed by $d^*(2380)$ hexaquarks. In a first study of such a
scenario it has been pointed out that stable $d^*(2380)$ condensates could
have formed in the early universe constituting even a candidate for dark
matter \cite{MB_DM}. 

\section{Acknowledgments}
\label{sec-acknowledge}

We acknowledge valuable discussions with M. Bashkanov, Stanley J. Brodsky,
Y. B. Dong, A. Gal, T. Goldman, Ch. Hanhart, Fei Huang, V. Kukulin, E. Oset,
M. Platonova, P. N. Shen, 
I. I. Strakovsky, Fan Wang, C. Wilkin, R. Workman and Z. Y. Zhang.
This work has been supported by DFG (CL 214/3-3). One of us  (H.~Cl.)
appreciates the support
by the Munich Institute for Astro- and Particle Physics
(MIAPP) which is funded by the Deutsche Forschungsgemeinschaft (DFG, German
Research Foundation) under Germany's Excellence Strategy -- EXC-2094 -
390783311.

\end{document}